\documentclass[5p,sort&compress]{elsarticle}

\usepackage{multirow,url,graphicx,siunitx,textcomp,helvet,array,makecell,hhline,lineno,hyperref,epstopdf,chemformula,booktabs,mathtools}
\usepackage[labelformat=simple,position=b]{subcaption}

\newcolumntype{M}[1]{>{\centering\arraybackslash}m{#1}}
\graphicspath{{Images/}}
\DeclareSIUnit\px{px}
\modulolinenumbers[5]

\journal{Measurement}

\begin{document}

\begin{frontmatter}
  \title{Thermal and dimensional evaluation of a test plate for assessing the measurement capability of a thermal imager within nuclear decommissioning storage}
\author{J McMillan\fnref{fn1}}
\author{M Hayes, R Hornby, S Korniliou, C Jones, D O'Connor, R Simpson and G Machin}
\address{National Physical Laboratory, Hampton Road, Teddington, TW11 0LW, UK}
\author{R Bernard and C Gallagher}	
\address{Sellafield Ltd., Seascale, Cumbria, CA20 1PG}
\fntext[fn1]{Corresponding author \texttt{jamie.mcmillan@npl.co.uk}}

\begin{abstract}
  In this laboratory-based study, a plate was designed, manufactured and then characterised thermally and dimensionally using a thermal imager. This plate comprised a range of known scratch, dent, thinning and pitting artefacts as mimics of possible surface anomalies, as well as an arrangement of higher emissivity targets. The thermal and dimensional characterisation of this plate facilitated surface temperature determination. This was verified through thermal models and successful defect identification of the scratch and pitting artefacts at temperatures from \SIrange{30}{170}{\celsius}.

These laboratory measurements demonstrated the feasibility of deploying in-situ thermal imaging to the thermal and dimensional characterisation of special nuclear material containers. Surface temperature determination demonstrated uncertainties from \SIrange{1.0}{6.8}{\celsius} (\(k = 2\)). The principle challenges inhibiting successful deployment are a lack of suitable emissivity data and a robust defect identification algorithm suited to both static and transient datasets.
\end{abstract}

\end{frontmatter}



\section{\bf Scope of Work} \label{sec:scope}

Ongoing condition monitoring and inspection of nuclear waste storage containers is a necessity for nuclear decommissioning activities internationally; the laboratory measurements detailed here seek to describe the suitability of using thermal imaging for the characterisation of these containers. Two container types used by Sellafield Ltd. are the Magnox and THORP special nuclear material containers. The non-contact characterisation for container condition monitoring and inspection using instrumentation translated along an inspection rail to view the container surfaces, would provide Sellafield Ltd. with critical information to support waste management decisions. Thermal imaging measures the apparent surface radiance temperature from objects within its field of view. Thermal imaging can be used to determine and evaluate surface temperature, surface photo-thermal properties and geometrical features.


\section{\bf Introduction} \label{sec:introduction}
Prior to the use of thermal imaging for container inspection, its measurement capability was evaluated in a controlled environment using a plate constructed from stainless steel 316L. The objective of the investigation was to identify the primary measurement capability and challenges of the hardware.

\subsection{Background} \label{subsec:background}
Special Nuclear Materials (SNM) are defined by the Atomic Energy Act of 1954 \cite{ref:atomic_energy_act} as plutonium or uranium and are usually stored in 316 stainless-steel containers. Common types of SNM containers are the Magnox and THORP. The former consists of two nested containers and the latter, three nested containers. The wall thickness of the external container is nominally \SI{1}{\milli\metre}. 

SNM containers will corrode over time and therefore it is important that degradation is monitored effectively. Accurate inspection of active SNM containers requires high quality quantitative data in order to demonstrate that they will reach their design lifetime. The detection of defects on the outside of containers in isolated locations is a technical challenge due to the harsh environment, reduced access and requirement for non-contact inspection. Special nuclear materials emit gamma rays and neutron radiation that may result in the production of heat which can be detected by a thermal imager.

\subsection{Thermal imaging for inspection of SNM containers} \label{subsec:non_destructive_testing}
Any object at a temperature above absolute zero emits infrared radiation and its distribution follows Planck’s law \cite{ref:thermal_radiation_heat_transfer}. The signal measured by each pixel of a thermal imager is dependent on the observed radiance emitted by the surface. Emissivity is a thermophysical property of a material and it describes how efficiently a surface can absorb and emit radiation. Emissivity is dependent on the wavelength, the temperature of the surface and both polar and azimuthal angles of measurement. The emissivity dependence of radiance temperature is shown in Eq.~\ref{eq:surface_temperature}, obtained from \cite{ref:thermal_radiation_heat_transfer}. A thermal imager measures the apparent radiance temperature of a surface captured within its Field Of View (FOV) and can observe surface features due to emissivity variations and real temperature variations on the surface.  

Two techniques to use thermal imagers include passive and active thermography; in this manuscript, passive thermal imaging was used to determine surface temperature of SNM containers. Active thermography was considered for the defect detection within this project but was not deployed.

Data processing algorithms have been used to improve the success of anomaly detection from thermal images and automate the inspection process \cite{ref:detection_subsurface_defects}; in particular, edge detection and thresholding algorithms were used to process raw thermal images \cite{ref:infrared_thermography_temperature_ndt}. The edge detection algorithm was used for identifying sharp discontinuities and thresholding was used for image segmentation. These binary images provided greater insight into the defect locations and the relative contrast corresponding to different defects. 

Throughout this paper, temperatures directly measured by the thermal imager are denoted as apparent radiance temperatures; these are not corrected for the non-unity emissivity. When an emissivity correction has been applied to this apparent radiance temperature it is described as the radiance temperature. This radiance temperature is in principle equivalent to the surface temperature measured using alternative thermometry methods such as contact thermometers and phosphor thermometers.

\subsection{Objective} \label{subsec:objective}

The objective of this research was to investigate the feasibility of thermal imagers being used to both measure surface temperature and identify features on the surface and sub-surface of a container, such as thinning, pitting, dents and scratches. If there is a temperature gradient along the container, it is necessary to be able to identify features based on the relative temperature as opposed to absolute temperature thresholds, this was used to validate the detection capability. To assess the dimensional measurement capability of the thermal imager, a metallic plate comprising a range of known scratch, dent, thinning and pitting artefacts was designed and constructed. Thermal characterisation of the plate and geometrical characterisation of the defects was performed. The thermal and dimensional characterisation of this plate surface temperature data were verified through thermal models and the defect identification method of the scratch and pitting artefacts were all tested at temperatures from \SIrange{30}{170}{\celsius}. Sub-surface detection of these artefacts using thermal imaging was also investigated.

The thermal imager used was traceable to the International Temperature Scale of 1990 (ITS-90) \cite{ref:npl_its90} and the uncertainties were evaluated according to the Guide to the Expression of Uncertainties \cite{ref:gum}.


\section{\bf Plate Design} \label{sec:simulant_artefact_design}
As introduced in Section~\ref{sec:introduction}, engineering a range of surface defects on a representative container and ensuring sufficient temperature monitoring would introduce too many uncontrolled variables. As an intermediary step, a stainless-steel target plate with known defects was designed and manufactured. The plate was designed to be positioned with the artefacts facing outwards or inwards to enable both surface and sub-surface defect detection to be evaluated.

\subsection{Design considerations} \label{subsec:design_considerations}
The primary considerations for low uncertainty surface temperature determination are: high rate of heat transfer from the heating element to radiating surface, low and known heat loss mechanisms, sufficient contact thermometer coverage, and known and low uncertainty photo-thermal properties for the radiating surface. 

\begin{figure}[t]
\centering
	\subcaptionbox{Cross-section of the assembly. \label{fig:simulant_assembly_scematic_cross_section}}
		{\includegraphics[width=0.5\textwidth,keepaspectratio]{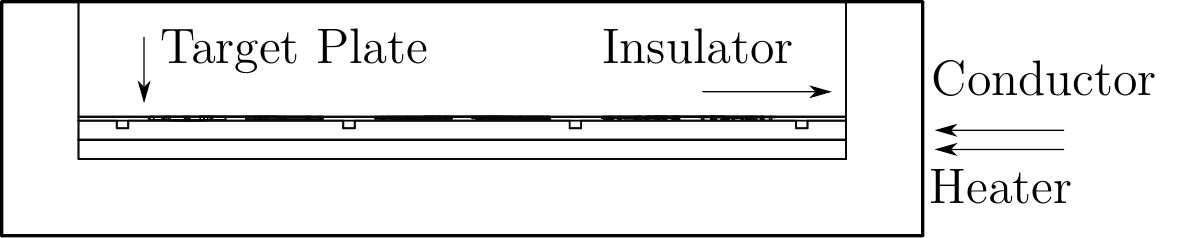}}
	\vskip\baselineskip
	\subcaptionbox{Lateral view of the target plate. \label{fig:simulant_assembly_scematic_target_plate}}
		{\includegraphics[width=0.5\textwidth,keepaspectratio]{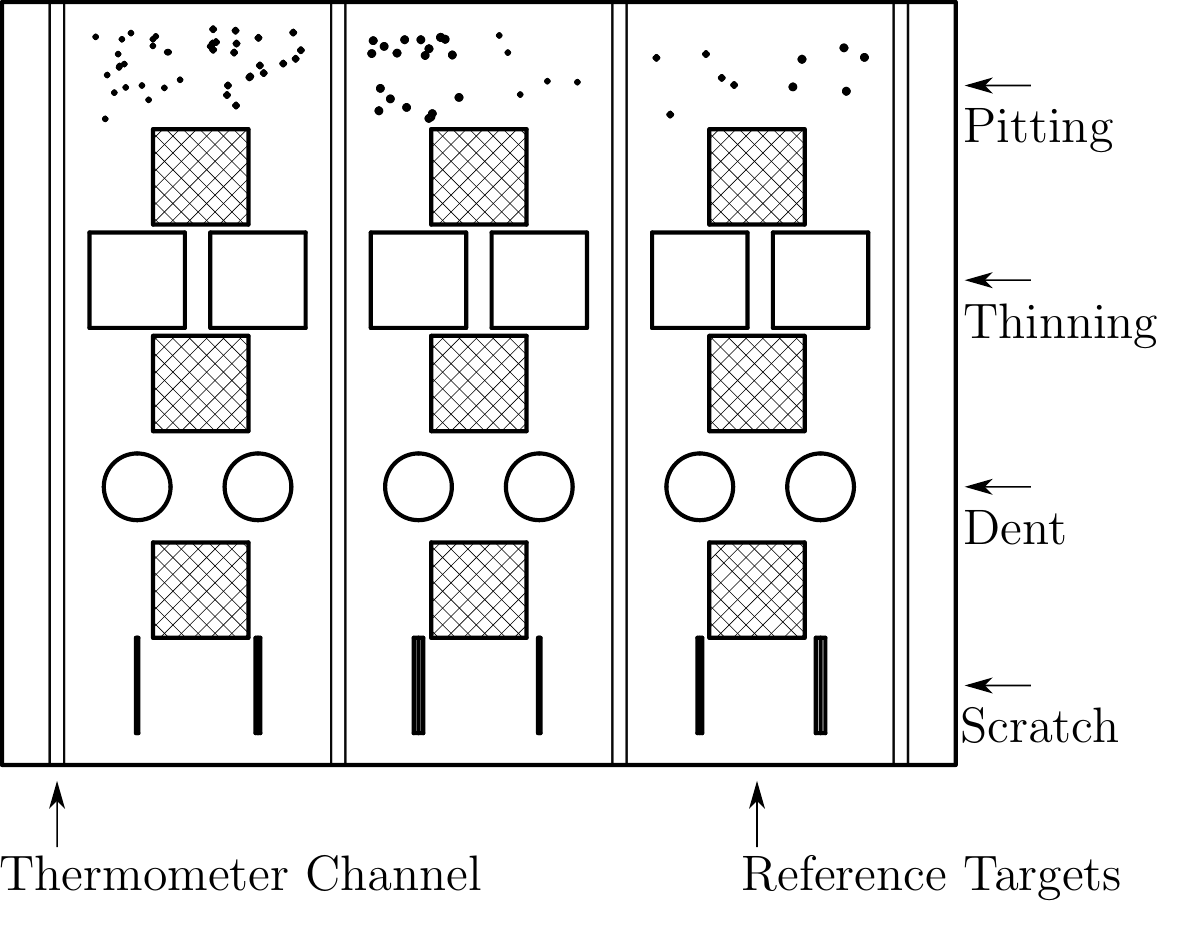}}
	\caption{The assembly designed for this investigation. (a) Cross-section of the assembly, from inside to out: target plate, thermal conductor plate, heater and housing. (b) Lateral view comprising the artefact zones and coated reference targets.}
	\label{fig:simulant_assembly_scematic}
\end{figure}

A schematic of a cross-section from the assembly can be seen in Fig~\ref{fig:simulant_assembly_scematic_cross_section}. For the highest heat conduction from the silicone rubber wire wound heater mat to the target plate, the two should be in direct contact with an appropriate pressure and thermal paste between them. To increase the heating uniformity for the target plate, an aluminium thermal conductor plate was placed in between. To reduce heat losses from the target-conductor-heater assembly, a low thermal conductivity housing was designed to surround it. This housing was designed to have a sufficiently thick wall in order to insulate the target-conductor-heater assembly and reduce heat transferred by natural convection to the surroundings.

An array of reference temperatures were determined through a set of thermocouples located between the target and thermal conductor plate. To avoid the thermocouples introducing an air gap between the two plates, a set of recessed thermometer channels were machined into the thermal conductor plate. To support the radiance temperature determination from the target plate, an array of coated regions was applied. The radiating surface of the target plate can be seen in Fig~\ref{fig:simulant_assembly_scematic_target_plate}.

The contact pressure was addressed in the final assembly with a pair of clamped blocks at either end and a thermal paste was omitted to reduce the contamination and subsequent degradation of the coated regions when transitioning between surface and sub-surface defect measurements. 

\subsection{Defect definition}\label{subsec:defect_definition}
The four types of artefact that will be investigated both in the surface and sub-surface configurations are: scratches, dents, thinning and pitting. The specifications for these artefacts are described below.

\begin{figure}[h!]
\centering
	\subcaptionbox{Scratch defect. \label{fig:simulant_scratch}}
{\includegraphics[width=0.23\textwidth,keepaspectratio]{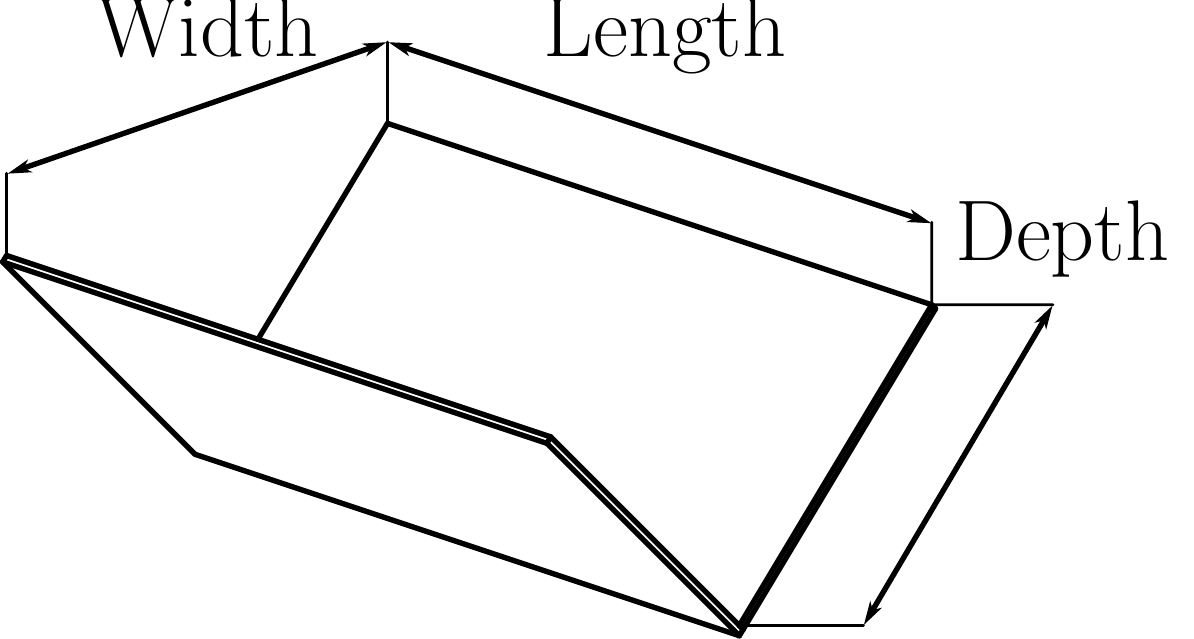}}
	\subcaptionbox{Dent defect. \label{fig:simulant_dent}}
		{\includegraphics[width=0.23\textwidth,keepaspectratio]{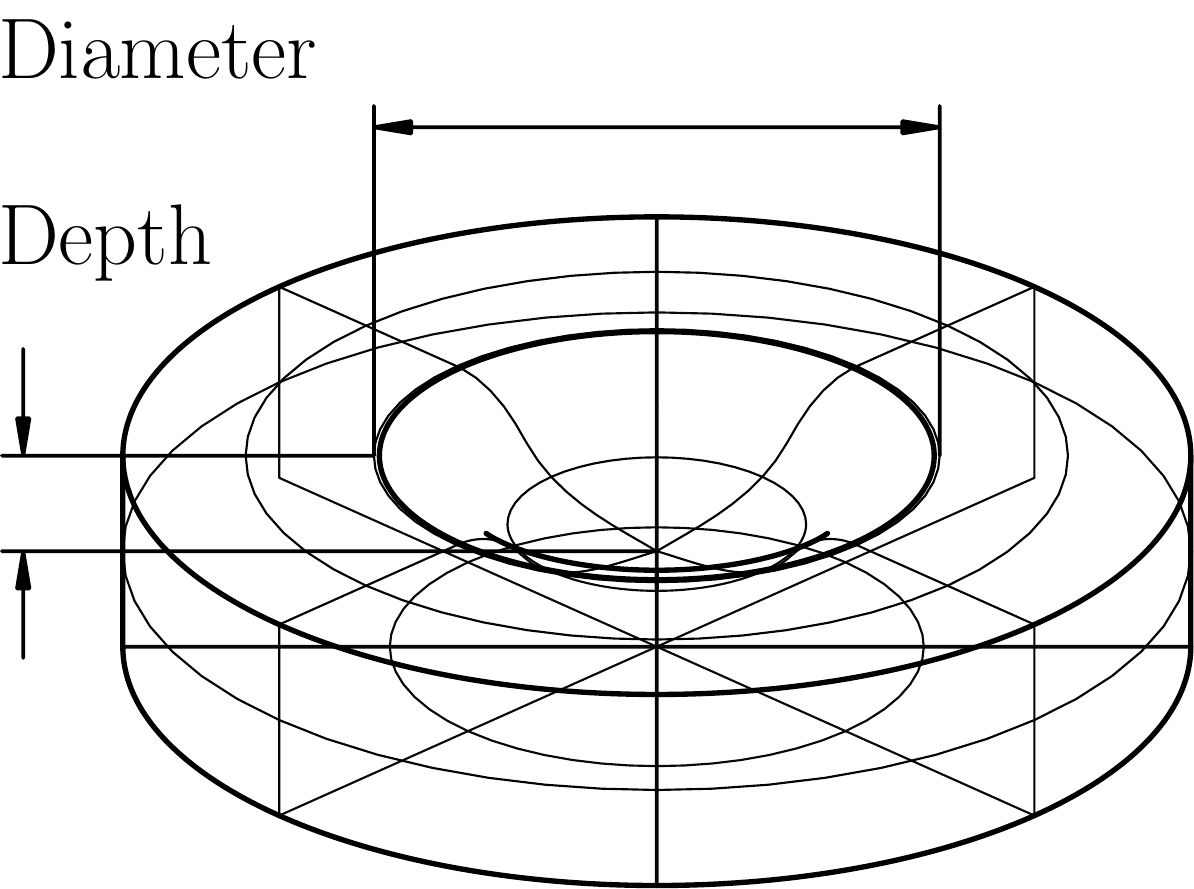}}
	\vskip\baselineskip
	\subcaptionbox{Thinning defect. \label{fig:simulant_thinning}}
		{\includegraphics[width=0.23\textwidth,keepaspectratio]{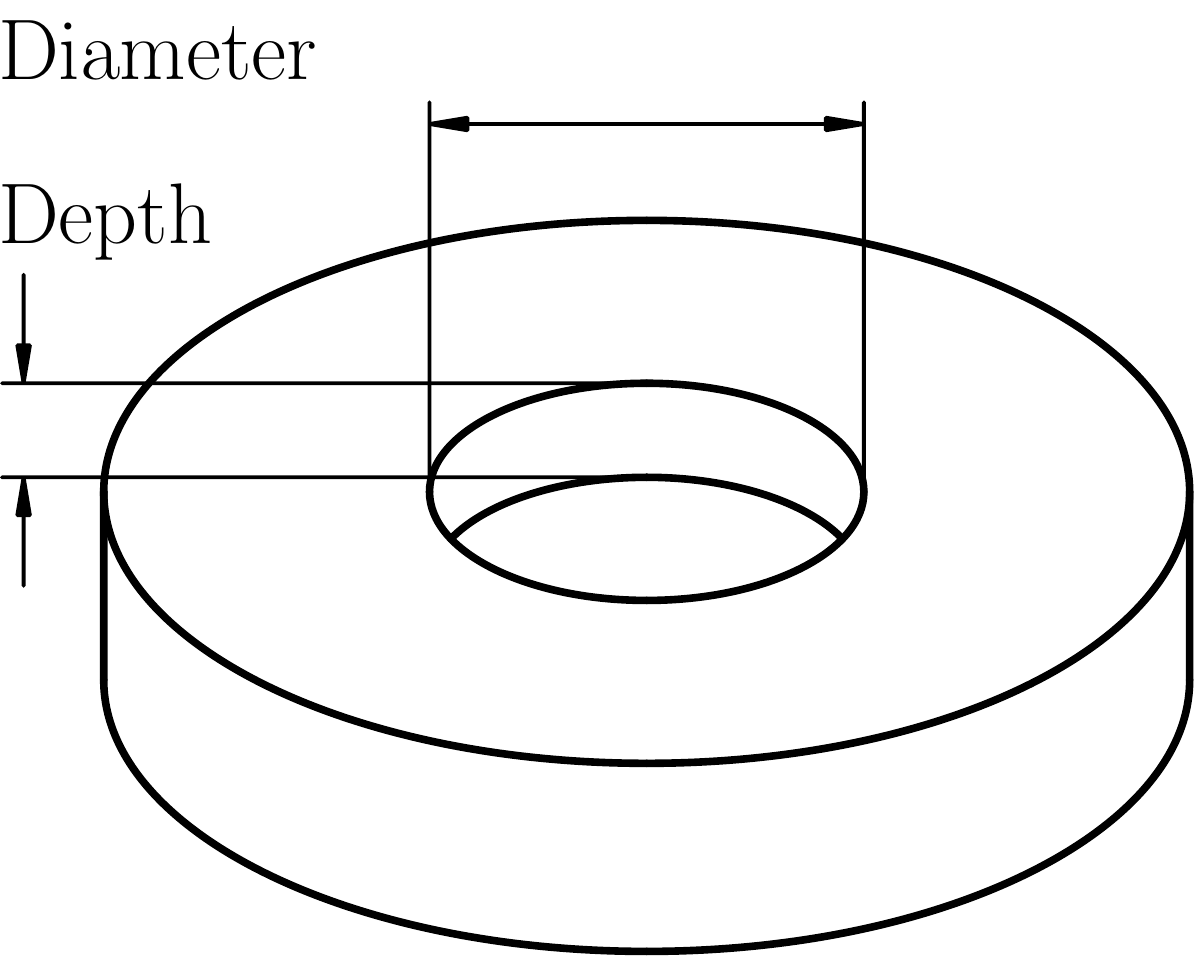}}
	\subcaptionbox{Pitting defect. \label{fig:simulant_pitting}}
		{\includegraphics[width=0.23\textwidth,keepaspectratio]{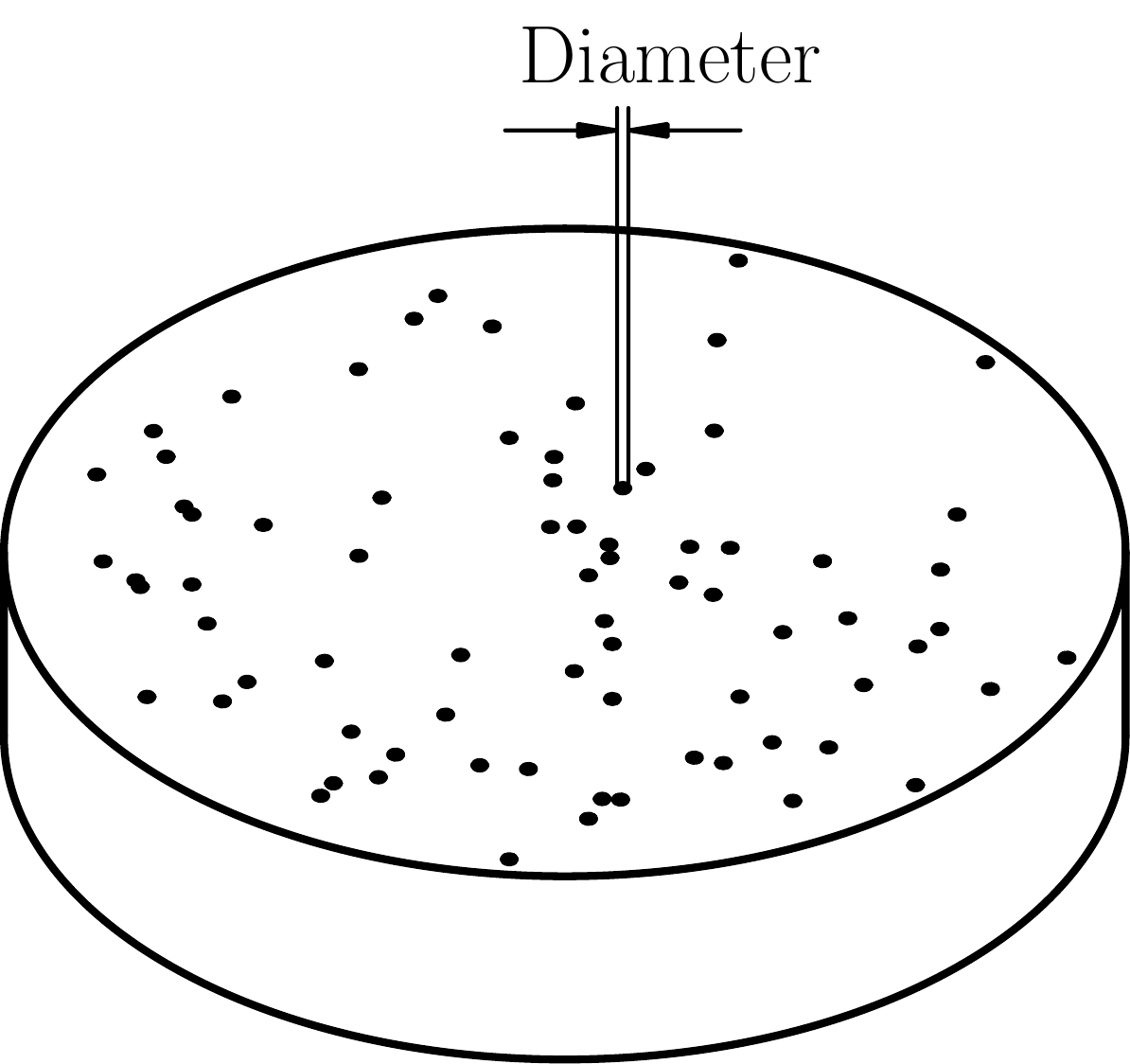}}
	\caption{The array of surface defects engineered and their geometrical definition.}
	\label{fig:defect_definition}
\end{figure}

A scratch will be defined as a vee-groove with nominally a \SI{45}{\degree} slope gradient. Scratch depth is the distance between the apex of the groove and the height of the neighbouring surface. This can be seen in Fig~\ref{fig:simulant_scratch}. The width is the distance between the two top edges of the vee. The length of the scratch is the end-to-end distance of the groove. The depth is the vertical distance between the apex and the top edge of the groove.

Dents are ellipsoidal impressions in a surface, this can be seen in Fig~\ref{fig:simulant_dent}. The depth of the impression is the distance between the top edge and the centre of the base. The diameter is the distance between two top edges where the impression begins.

Surface thinning is a recess into a surface, this can be seen in Fig~\ref{fig:simulant_thinning}. The depth of the recess is the distance between the top edge and the base. The diameter is the distance between the two top edges where the recess begins.

Surface pitting is a random array of cylindrical voids of a specified diameter and depth, this can be seen in Fig~\ref{fig:simulant_pitting}. The diameter of the void distance between the two top edges where the void begins. The depth of the voids is defined and identical. The arrangement of the voids is arbitrary but is localised to the artefact region.

\subsection{Defect manufacturing} \label{subsec:defect_manufacturing}
The housing, insulator, thermal conductor and heater components were manufactured by external manufacturers. The insulator was machined from a cement based high-temperature insulation board Sindanyo H91, which has a thermal conductivity of \SI{0.5}{\watt\per\metre\per\kelvin} and a wall thickness of \SI{20}{\milli\metre}. The thermal conductor plate is \SI{5}{\milli\metre} thick aluminium with \SI{4}{\milli\metre} by \SI{2}{\milli\metre} profile thermometer channels. 

The target plate was manufactured from stainless-steel 316L, its geometry was \SI{1}{\milli\metre} thick, \SI{200}{\milli\metre} long and \SI{160}{\milli\metre} wide. The artefact zones were manufactured both internally and externally. The pitting was manufactured using centre drills, the scratches and thinning were manufactured using a \SI{12}{\milli\metre} carbide slot drill. The dents were manufactured using a range of ball bearings mounted within an Avery 7110 compression machine that were loaded up to \SI{30}{\kilo\newton}. From left to right (in Fig~\ref{fig:simulant_assembly_scematic_target_plate}) the dents were manufactured using a \SI{10}{\milli\metre} ball bearing under \SI{10}{\kilo\newton}, \SI{20}{\kilo\newton} and \SI{30}{\kilo\newton} load, and a \SI{40}{\milli\metre} ball bearing under \SI{10}{\kilo\newton}, \SI{20}{\kilo\newton} and \SI{30}{\kilo\newton} load respectively. Following the artefact manufacturing, the \SI{20}{\milli\metre} square reference targets were coated with Senotherm Ofen-spray schwarz 17-1644-702338 paint in order to ensure a higher emissivity region for surface radiance temperature measurement. From previous experience and similar coatings, it is anticipated that the emissivity of this material is close to 0.85 \cite{ref:senotherm_emissivity}. The completed artefact can be seen in Fig~\ref{fig:simulant_plate_image}.


\section{\bf Method} \label{sec:method}
To support the thermal imager measurements of the plate, a thermal model was designed and evaluated, the temperature response of the thermal imager was calibrated and the dimensional topography of the plate independently measured. 

Following these activities, the calibrated thermal imager was set up to view the appropriate surface of the plate for the thermal and geometrical evaluation. Thermal images were captured in two minute sequences using each appropriate integration time. The thermocouple measurements were captured simultaneously.

\subsection{Thermal model configuration} \label{subsec:thermal_model}
\begin{figure}[h]
		\centering
\includegraphics[width=0.5\textwidth,keepaspectratio]{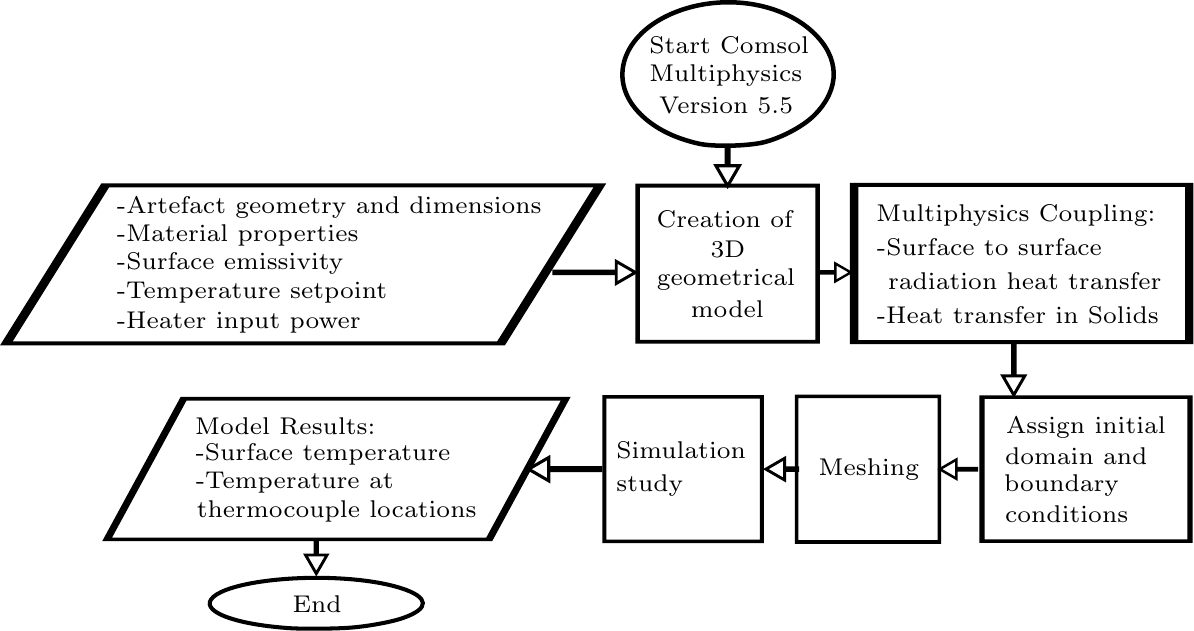}
		\caption{Flow chart of modelling steps using COMSOL.}
		\label{fig:model_logic_flow}
\end{figure}

In order to model heat transfer in the plate assembly, the heat transfer in solids interface was coupled with the surface to surface radiation interface in COMSOL Multiphysics V5.5. The solution was derived using finite element analysis. The computational problem was a 3D coupled conduction, convection and radiation heat transfer problem. In order to model this heat transfer process, both conduction in the solid, and natural convection in the surrounding fluid were taken into account. For natural convection, known correlations for heat transfer coefficients were used. The input, output and model flow process is described in a flow chart in Fig~\ref{fig:model_logic_flow}.

\begin{figure}[h]
		\centering
		\includegraphics[width=0.5\textwidth,keepaspectratio]{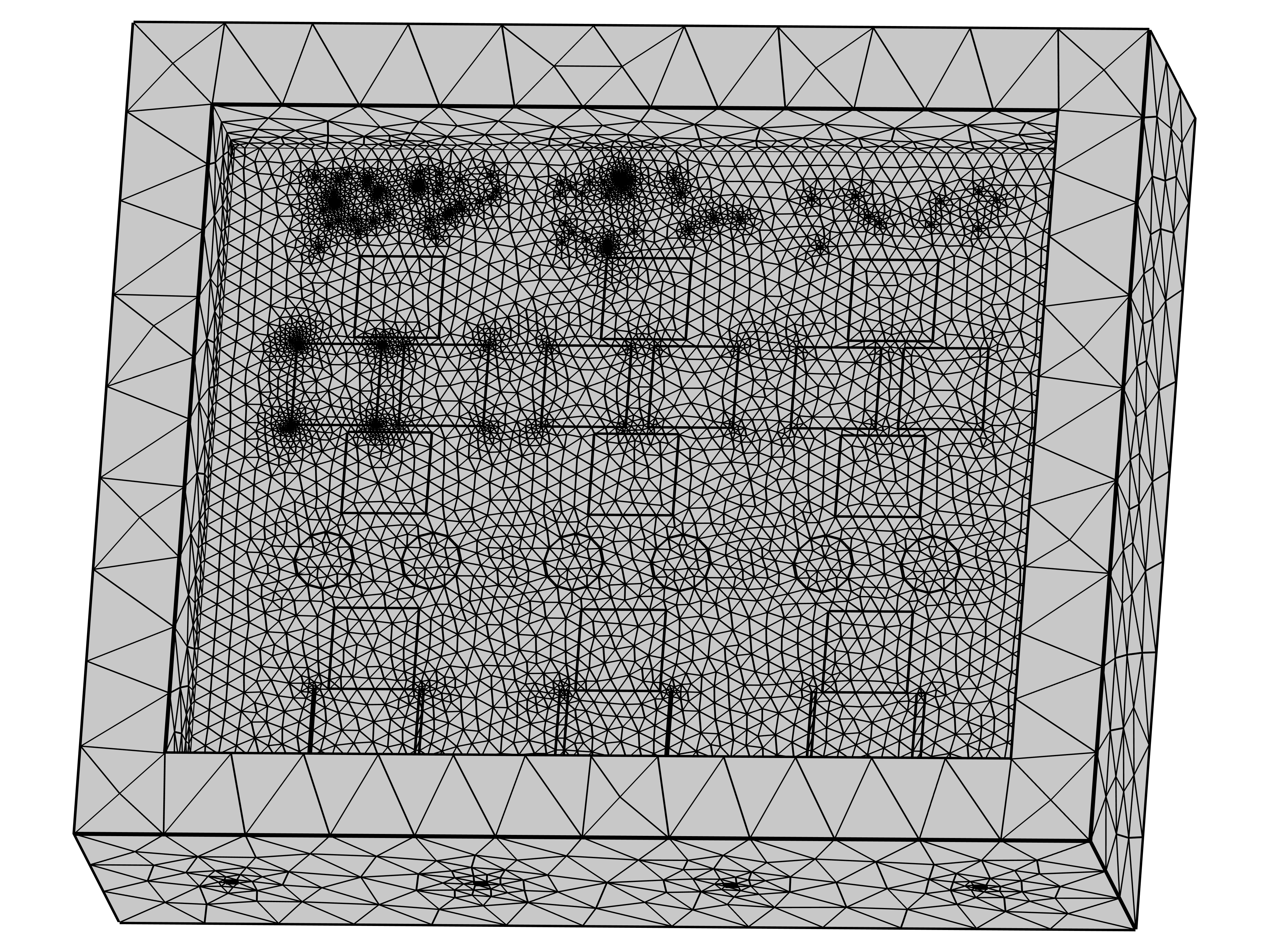}
		\caption{3D mesh of target-conductor-heater-insulator assembly model produced by COMSOL. }
		\label{fig:simulant_plate_mesh}
\end{figure}

The physical model of the plate assembly consists of the acrylic heater mat, the aluminium thermal conductor plate, the stainless-steel reference target plate and the H91 insulator. The 3D mesh solved with the finite element method is shown in Fig~\ref{fig:simulant_plate_mesh}. The thermal boundary conditions included input power from \SIrange{5.3}{167.5}{\watt} and temperature setpoints from \SIrange{30}{170}{\celsius}. For the surface to surface radiation heat transfer, an emissivity of \num{0.3} was used, obtained from literature as a mean value between \num{0.25} and \num{0.35} \cite{ref:emissivity_ss316l}.

\subsection{Thermal characterisation of imager} \label{subsec:thermal_characterisation}

\begin{figure}[h]
\centering
\includegraphics[width=0.5\textwidth,keepaspectratio]{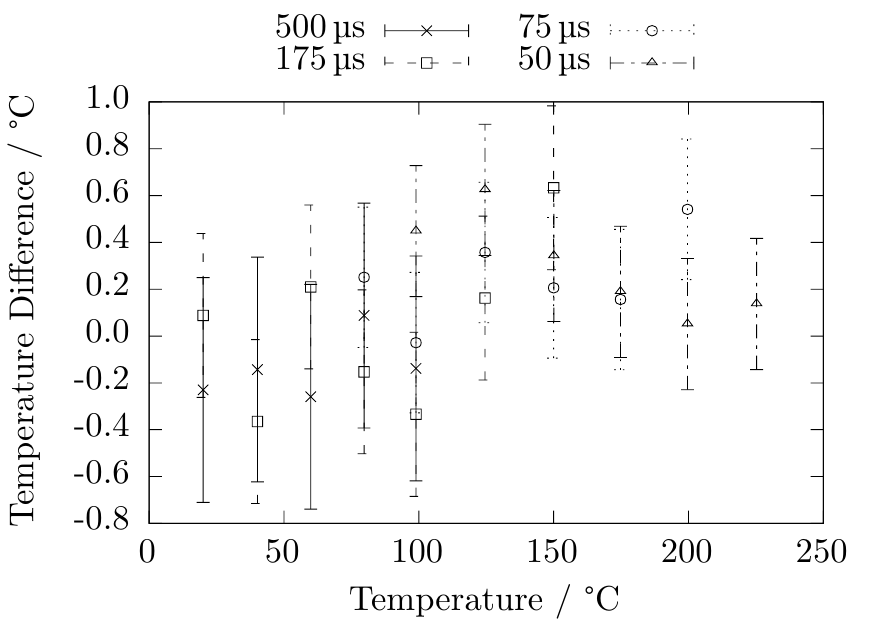}
\caption{The validation measurements of the calibrated thermal imager. These measurements show the temperature difference between ITS-90 and the apparent radiance temperature from the thermal imager. The error bars indicate the expanded uncertainty with a coverage probability of \SI{95}{\percent}.}
\label{fig:thermal_imager_validation}
\end{figure}

A cooled medium-wave infrared thermal imager was used throughout this project, an InfraTec ImagerIR 8300 using a \SI{25}{\milli\metre} lens (spectral range from \SIrange{2.0}{5.7}{\micro\metre}). This was calibrated against a series of reference cavity blackbody sources from \SIrange{20}{225}{\celsius} as detailed in \cite{ref:npl_cavity_references}. The suitability of this calibration across the temperature range for four integration times can be seen in Fig~\ref{fig:thermal_imager_validation} by comparison of the calibrated apparent radiance temperature against ITS-90. In addition to this, additional temperature response checks accounting for the size-of-source effect \cite{ref:sse} and distance variation were carried out. 

\subsection{Dimensional characterisation of plate} \label{subsec:dimensional_characterisation}
To understand the surface defect detection capability of the thermal imager, its dimensional measurement characteristics needed to be calibrated against a suitable reference. This was achieved by using a 3D optical profilometer to measure the dimensions of the plate features, which were then compared with thermal imager measurements of the same features.

The 3D optical profilometer used was an NPL-developed hybrid 2D/3D measurement platform \cite{ref:HDRPlatform} configured as a large-area Chromatic Confocal Microscope (CCM), capable of measuring the complete areal topography of the plate in one operation. The measured topographies were then used to calculate individual plate feature dimensions.

The CCM consists of a chromatic confocal point probe sensor with a \SI{12}{\milli\metre} measuring range mounted vertically (\(z\)) above a lateral (\(x,y\)) sample translation system. The plate was moved in a raster scan to build up a 2D grid of surface heights measured using the chromatic confocal principle \cite{ref:ChrConf}. 

\subsection{Experimental setup} \label{subsec:experimental_setup}
\begin{figure}[h!]
\centering
\includegraphics[width=0.5\textwidth,keepaspectratio]{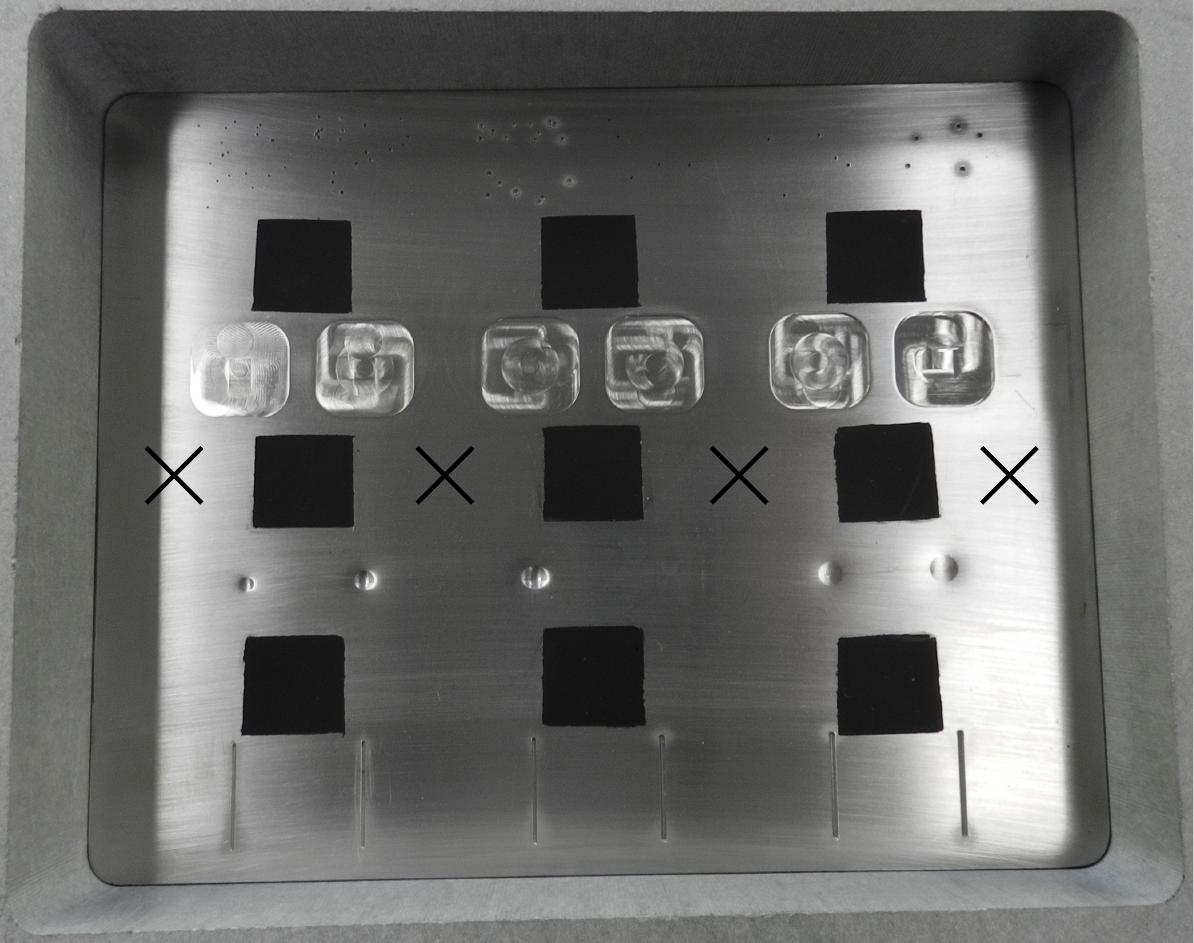}
\caption{The target plate can be seen, the thermocouple locations are indicated by crosses.}
\label{fig:simulant_plate_image}
\end{figure}

\begin{figure}[h!]
\centering
\includegraphics[width=0.3\textwidth,keepaspectratio]{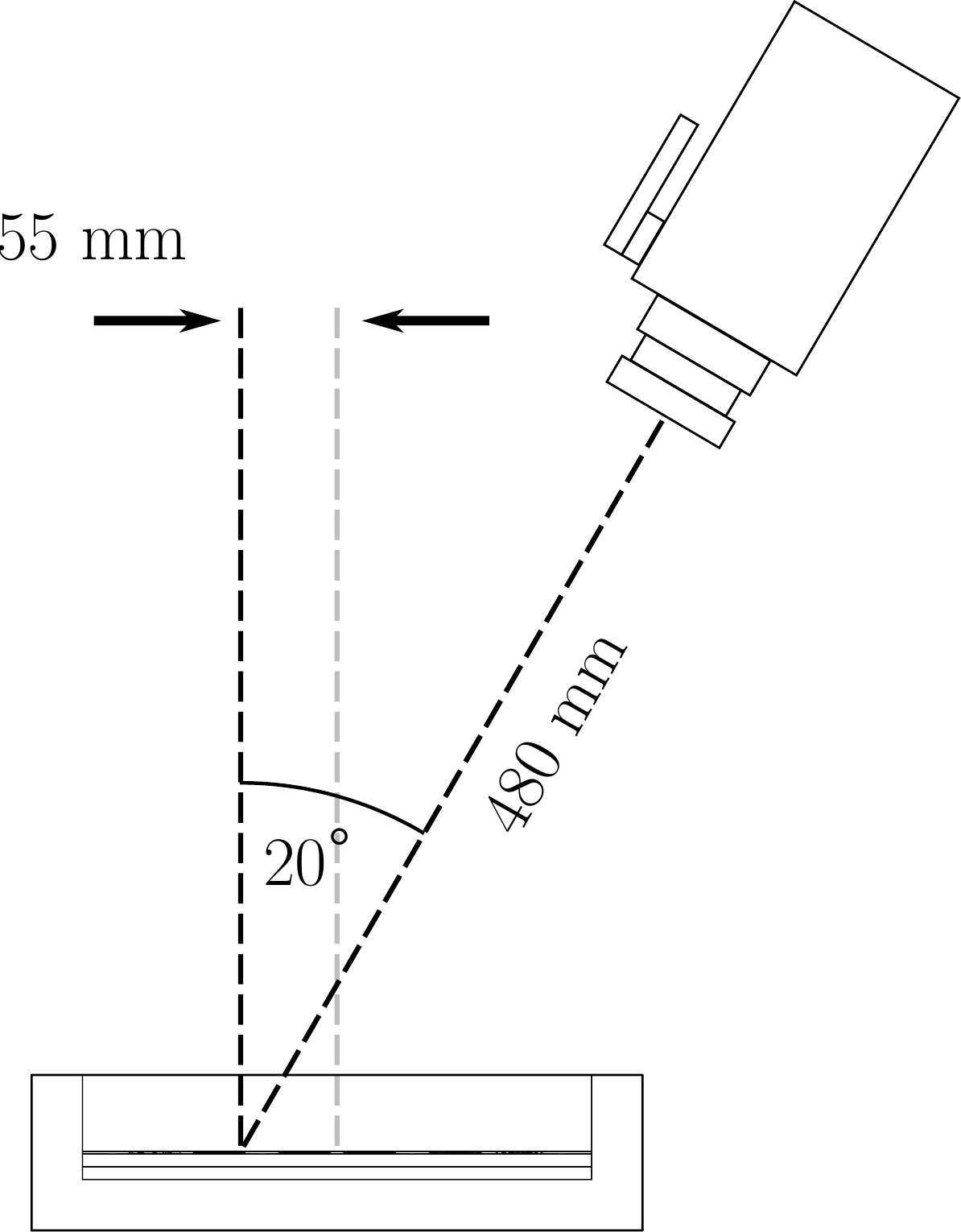}
\caption{Schematic diagram of the thermal imager observing the plate. The imager was \SI{20}{\degree} from the surface normal and there was nominally \SI{480}{\milli\metre} between the plate to the front of the lens. The distance from the centre of the plate to the intersection of the imager optical axis was nominally \SI{55}{\milli\metre}.}
\label{fig:simulant_imager_configuration}
\end{figure}

The plate was set up according to the schematic in Fig~\ref{fig:simulant_assembly_scematic}. One control thermocouple was mounted to the heater surface (between the heater and thermal conductor); one monitoring thermocouple was located within the centre of each thermometer channel, these were each class one type K thermocouples. The monitoring thermocouples were connected to a Fluke 1586A Super-DAQ precision temperature scanner. Each scanner channel was configured as a type K thermocouple and used the same internal reference point. The plate and thermocouple locations can be seen in Fig~\ref{fig:simulant_plate_image}. The thermal imager was mounted \SI{20}{\degree} from the surface normal of the plate and at a distance of \SI{480}{\milli\metre} from the plate to the front of the lens, this can be seen in Fig~\ref{fig:simulant_imager_configuration}. At this angle, the distance from centre of the lens to both the far and near edges on the plate was estimated to range from \SIrange{490}{450}{\milli\metre}.

To improve conductive heat transfer from the heater through to the target plate, a pair of clamps was used with anodised aluminium blocks to apply pressure to the top of the target plate. During the measurements, a shroud was placed over the imager and plate to reduce background reflection variations and natural convection.

For the sub-surface plate measurements, the same setup was used, albeit with the target plate reversed such that the pitting artefacts remained the farthest artefact from the thermal imager, then the blocks were replaced.

\subsection{Defect detection} \label{subsec:defect_detection}

The feasibility of detecting defects on the surface of the plate was investigated.

There have been a number of visual-based approaches to scratch detection in the literature, such as using morphological operations, convoluted neural networks and image processing techniques \cite{ref:scratch_detection_1,ref:scratch_detection_2,ref:scratch_detection_3}. However, in this body of work scratches, were identified based on the edge detected image through the use of thermal image data.

After applying a Canny edge detector, a set of morphological operations were applied to close the binary mask, remove any small objects and fill in holes. Contours in the image were then identified. Given that scratches are generally thin and highly directional, the contours were filtered by their inertia ratio (how elongated the shape is) and by the thickness of the shape.

Pitting on the surface was identified as small, circular clusters of pixels in the edge detected image.

Dents and regions of thinning were not able to be identified due to a lack of thermal contrast caused by these defects.


\section{\bf Uncertainties} \label{sec:uncertainties}
Each individual contribution to total measurement uncertainty can be attributed to each instrument used for measurand determination, in this case the thermocouple, the thermal imager and the confocal microscope. These components have been evaluated with respect to their uncertainty value, probability distribution and sensitivity to the measurand. The standard uncertainty is then combined in quadrature to determine an expanded uncertainty. The uncertainty budgets have been considered according to the Guide to Uncertainty in Measurement (GUM) \cite{ref:gum}.

\subsection{Defect dimensional uncertainty} \label{subsec:uncertainty_defect}
Uncertainties in CCM measurements were estimated by following the guidance in ISO 25178-602:2010 \cite{ref:bsi_ChrConf}, using metrological characteristics which describe the measurement uncertainty components in \(x\), \(y\) and \(z\). The \(z\) components include linearity deviation, amplification coefficient, and noise in the output of the CCM. Further contributions to \(x\), \(y\) and \(z\) uncertainties came from: deviation from linear, rotation-free motion of the \(x\) and \(y\) motion stages in the form of Abbe errors and flatness, straightness and linearity deviations; optical properties including limited resolution and refractive index variations during measurement; and calibration data from manufacturer data sheets for the chromatic confocal probe and motion stages. 

{\em Lateral dimensions}: The uncertainty in the reported lateral dimensions is dominated by the sampling density of the measurements. Smaller contributing terms include the lateral optical resolution of the CCM probe (related to optical spot size) and the uncertainty in lateral motion generated by the sample translation system. The sample spacing of \SI{0.1}{\milli\metre} used in this study resulted in a typical uncertainty of \SI{0.13}{\milli\metre} (\(k = 2, n = 15\) points in each edge) in a scratch width estimate. This uncertainty is appropriate for the provision of plate reference measurements for characterisation of the thermal imager, which has a lateral dimensional resolution greater than \SI{0.3}{\milli\metre}. 

{\em Height dimensions}: All of the feature heights were calculated as a difference between the mean heights of two surfaces, such as the top and bottom surfaces of the thinning features. Uncertainties were calculated as a combination of the vertical uncertainty components of the CCM probe and the standard uncertainty of the mean of both the upper and lower feature surfaces, which includes deviations from the nominal feature geometry in the uncertainty. For example, the uncertainty in the depth of a typical thinning feature is \SI{0.01}{\milli\metre} (\(k = 2, n = 100\) points in the upper and lower surfaces).

\subsection{Thermocouple uncertainty} \label{subsec:uncertainty_thermocouple}
{\em Standard tolerance}: \cite{ref:bsi_thermocouples} describes the standard tolerance of a type K thermocouple to be \SI{1.5}{\celsius} or \(0.004 \times T \mathrm{[\SI{}{\celsius}]}\), whichever is larger. Given that the measurements in this project consider temperatures below \SI{375}{\celsius} this component is consistently a rectangular distribution throughout each budget.

{\em Thermometer bridge accuracy}: the manufacturer stated accuracy for temperature measurement using thermocouples is \SI{0.2}{\celsius}. This is applied as a normal distribution component in each budget.

{\em Measurement stability}: the mean standard deviation of the four thermocouples over the measurement at each temperature setpoint was calculated. This corresponds to the temporal stability of the complete thermocouple assembly but does not consider the spatial uniformity measured by the thermocouple arrays. The respective values were applied as a normal distribution component in each budget.

\subsection{Thermal imager uncertainty} \label{subsec:uncertainty_thermal}
{\em Calibration}: this includes components for: the calibration of the reference source, stability of the reference source, stability of the imager, resolution of the imager, drift of the reference source and the residual of the calibration fit. This varied between the four integration times but was below \SI{0.5}{\celsius} (\(k = 2.3\)) and the appropriate values were applied as a normal distribution to each budget.

{\em Housing temperature stability}: during each measurement, the housing temperature of the thermal imager was recorded and the standard deviation during each respective dataset was applied to the budgets as a normal distribution. The relationship between measured temperature and the housing temperature is not one-to-one and is understood to have less of an effect than this and so this an over estimation of this component. The effect from the offset of the housing temperature in application from that during calibration is known to impact the measurement but measurement data was not available to support its evaluation.

{\em Size-of-source effect}: the size-of-source effect was evaluated and the repeatability of \SI{0.18}{\celsius} was scaled from \SI{170}{\celsius} to the respective temperatures measured throughout each measurement. These values were applied as a rectangular distribution component in each budget. 

{\em Distance effect}: the distance effect was evaluated and the repeatability of \SI{0.29}{\celsius} was scaled from \SI{170}{\celsius} to the respective temperatures measured throughout each measurement. These values were applied as a rectangular distribution component in each budget. 

{\em ROI uniformity}: during each measurement set, the mean spatial standard deviation of apparent radiance temperature measured by the thermal imager at the coated Regions Of Interest (ROI) was evaluated. This was considered as a normal distribution component in each budget. The sensitivity function for this component is described by Eq.~\ref{eq:apparent_temperature_sensitivity}.

{\em Emissivity measurement}: for the coated regions on the surface, the emissivity value of \num{0.85} was approximated from the reference measurement \cite{ref:senotherm_emissivity}. This value was measured by the authors of the referenced article using instrumentation with an uncertainty reported to be \num{0.04} (no confidence factor was reported, so is assumed to be \num{1}). This value was applied as a normal distribution to each measurement set. The sensitivity function for this component is described by Eq.~\ref{eq:emissivity_sensitivity}.

{\em Emissivity interpretation}: for the coated regions on the surface, the emissivity value of \num{0.85} was approximated from the reference measurement \cite{ref:senotherm_emissivity}. This approximation was evaluated based on emissivity data at three different temperatures and over the spectral range from \SIrange{2.7}{5.0}{\micro\metre}. The uncertainty incurred from estimating from this data is \num{0.05}. This value was applied as a normal distribution to each measurement set. The sensitivity function for this component is described by Eq.~\ref{eq:emissivity_sensitivity}.

{\em Ambient temperature measurement}: due to the non-unity emissivity of the surfaces measured, the uncertainty of the ambient temperature has an influence on the calculated surface temperature. The measurement uncertainty of the calibrated hygrometer was \SI{0.1}{\celsius} (\(k = 1\)) across the appropriate temperature range and the variation of room temperature during each measurement dataset remained below this value. This value was applied as a normal distribution throughout each dataset using the sensitivity function described by Eq.~\ref{eq:ambient_temperature_sensitivity}.

Considering the components introduced in Section~\ref{subsec:uncertainty_thermocouple} and Section~\ref{subsec:uncertainty_thermal}, an example complete budget is detailed for the temperature determination methods used: thermocouple and thermal imager.

\begin{table}[ht]
		\centering 
		\caption{An example complete uncertainty budget for temperature determination at the \SI{170}{\celsius} setpoint with the surface plate and \SI{50}{\micro\second} integration time. Each component is attributed to the respective instrument (thermocouple or thermal imager). The uncertainty, \(u\), is reported in degrees Celsius except for the emissivity components, the necessary divisor and sensitivity and presented as the standard uncertainty, \(U\). These were then combined in quadrature and multiplied by the coverage factor.}
		\begin{tabular}{ M{7.0em} M{3.0em} M{3.0em} M{3.0em} M{3.0em} }
		\Xhline{1.0pt}
		Source &  \(u\) & Div. & Sens. & \(U\) / \SI{}{\celsius} \\
		\cmidrule(lr){1-5}
				\multicolumn{5}{c}{Thermocouple} \\
		\cmidrule(lr){1-5}
				Tolerance & 1.50 & 1.73 & 1.00 & 0.87 \\
				Thermometry bridge accuracy & 0.20 & 1.00 & 1.00 & 0.20 \\
				Stability & 0.03 & 1.00 & 1.00 & 0.03 \\
				\cmidrule(lr){2-5}
				\multicolumn{4}{r}{Expanded uncertainty (\(k = 2\))} & {\bf 1.8} \\
		\cmidrule(lr){1-5}
				\multicolumn{5}{c}{Thermal Imager} \\
		\cmidrule(lr){1-5}
				Calibration & 0.30 & 2.20 & 1.00 & 0.14 \\
				Housing temperature & 0.03 & 1.00 & 1.00 & 0.03 \\
				Size-of-source & 0.18 & 1.73 & 1.00 & 0.10 \\
				Distance & 0.29 & 1.73 & 1.00 & 0.17 \\
				ROI non-uniformity & 1.06 & 1.00 & 1.04 & 1.10 \\
				Emissivity measurement & 0.04 & 1.00 & 50.03 & 2.00 \\
				Emissivity interpretation & 0.05 & 1.00 & 50.03 & 2.50 \\
				Ambient temperature & 0.10 & 1.00 & 0.01 & 0.00 \\
				\cmidrule(lr){2-5}
				\multicolumn{4}{r}{Expanded uncertainty (\(k = 2\))} & {\bf 6.8} \\
		\Xhline{1.0pt}
		\end{tabular} 
		\label{tab:uncertainty}
\end{table}

Tab.~\ref{tab:uncertainty} describes the complete uncertainty budget describing the measurements at the \SI{170}{\celsius} temperature setpoint for the surface plate using the \SI{50}{\micro\second} thermal imager integration time. Temperature measurement from the thermocouple is representative of that within the thermometer channel and does not account for any thermal heat transfer between the channel and radiating surface. The uncertainty value \(u\) is reported as degrees Celsius unless otherwise stated (for the emissivity components). The divisor describes the probability distribution of the component, either a normal or rectangular distribution. The sensitivity is \num{1.00} for surface temperature components but requires conversion from other units to a standard uncertainty in degrees Celsius. The respective component standard uncertainties were combined in quadrature and the combined uncertainty was multiplied by \num{2.0} for a \SI{95}{\percent} confidence interval.


\section{\bf Results} \label{sec:results}
Following two measurement sequences of the plate in the surface and sub-surface arrangements, the defect identification and surface temperature determination results were evaluated and are presented below. The surface temperature was calculated throughout this project using the apparent radiance temperature (from the calibrated thermal imager), the ambient temperature (from the calibrated hygrometer), the instrument spectral range midpoint (\SI{3.85}{\micro\metre}) and the estimated emissivity for the appropriate surface, using Eq.~\ref{eq:surface_temperature}.

\begin{figure}[h!]
\centering
	\subcaptionbox{Surface configuration. \label{fig:simulant_surface_170degC_temperature_distribution}}
		{\includegraphics[width=0.48\textwidth,keepaspectratio]{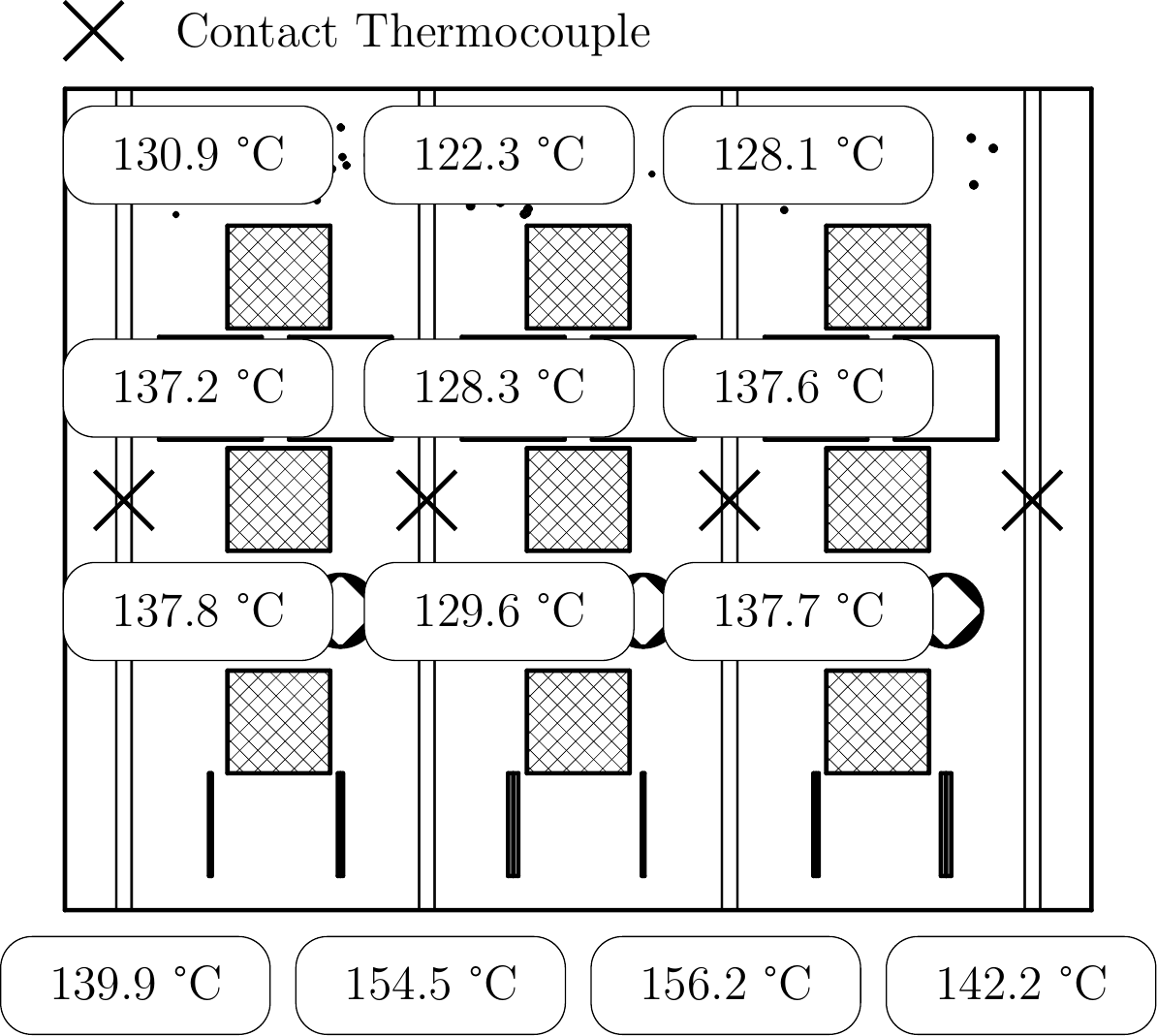}}\\
	\subcaptionbox{Sub-surface configuration. \label{fig:simulant_subsurface_170degC_temperature_distribution}}
		{\includegraphics[width=0.48\textwidth,keepaspectratio]{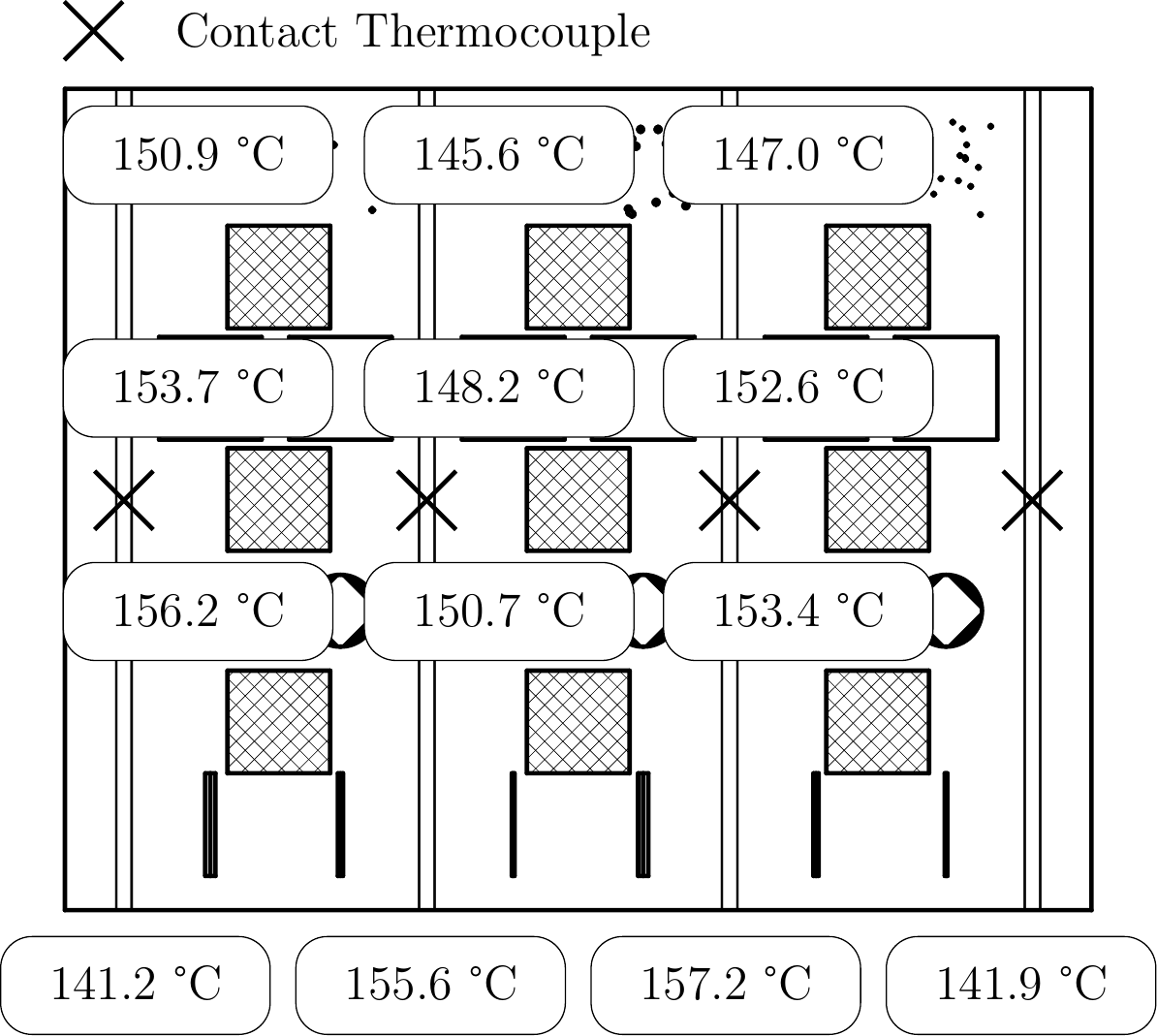}}
		\caption{The measured temperature distribution of the plate during measurements at \SI{170}{\celsius}. The four measurements at the bottom correspond to the respective thermocouples indicated by the crosses. The nine central values denote the emissivity corrected radiance temperatures of the adjacent regions. During the sub-surface measurements the same array locations to the surface configuration coated regions were evaluated. Note that the artefacts are shown for context but both coated regions and artefacts were present on the rear surface from view in the sub-surface measurements.}
	\label{fig:results_full_plate}
\end{figure}

The radiance and thermocouple temperatures for the surface and sub-surface plate configurations at a heater setpoint temperature of \SI{170}{\celsius} can be seen in Fig.~\ref{fig:results_full_plate}. Here the radiance temperature for the coated regions (hatched squares) are evaluated as the mean from the image series and over the appropriate integration times; the thermocouple temperatures were similarly evaluated over the measurement sequence and their location is indicated by the cross.

\begin{figure*}[h!]
\centering
\includegraphics[width=0.92\textwidth,keepaspectratio]{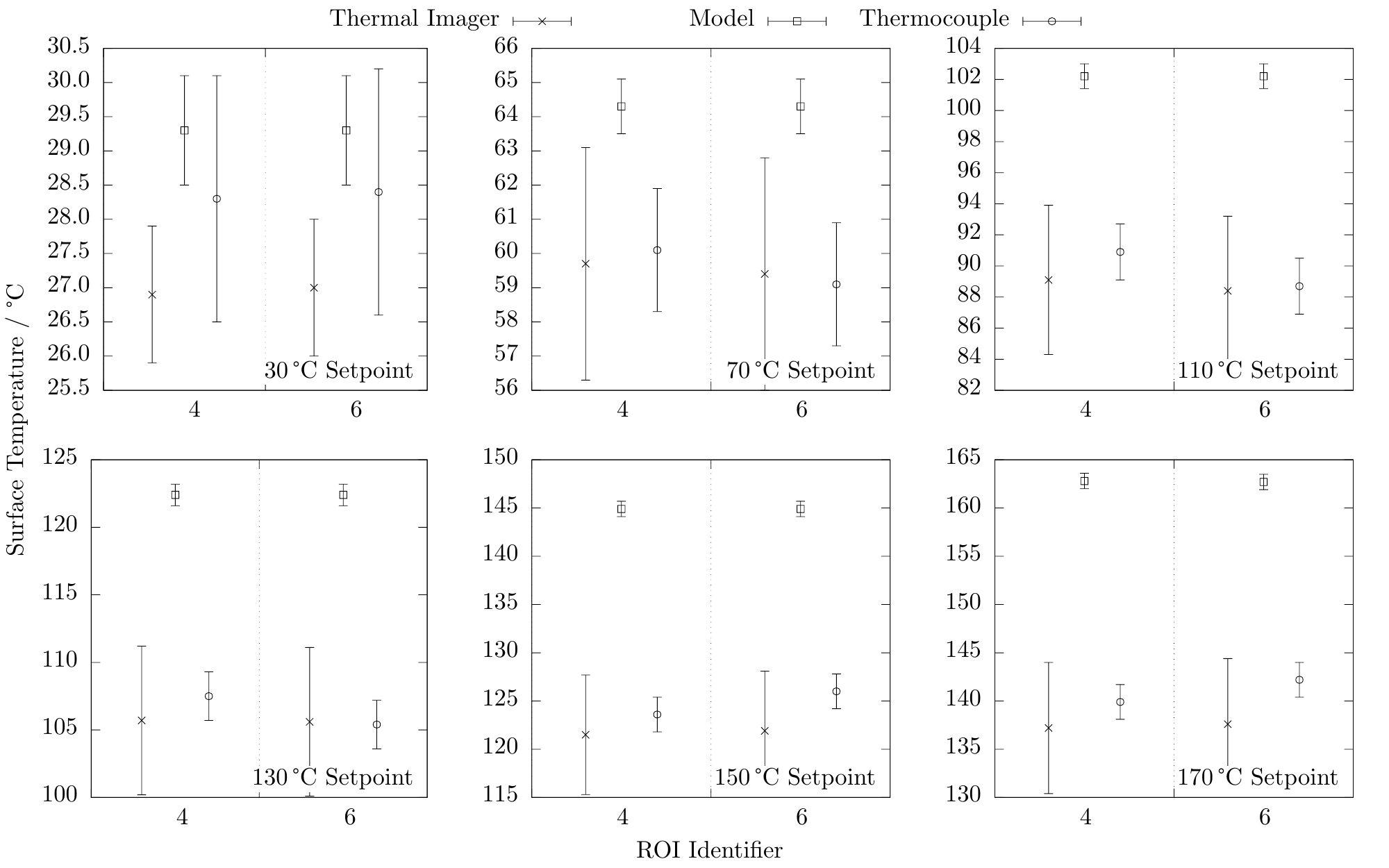}
\caption{A comparison between the surface temperature determinations of the coated regions using the thermal imager, thermocouples and the model with their equivalent uncertainties (\(k = 2\)) at each temperature setpoint.}
\label{fig:simulant_surface_artefact_test_plan_uncertainty_comparison}
\end{figure*}

\begin{figure*}[h!]
\centering
\includegraphics[width=0.92\textwidth,keepaspectratio]{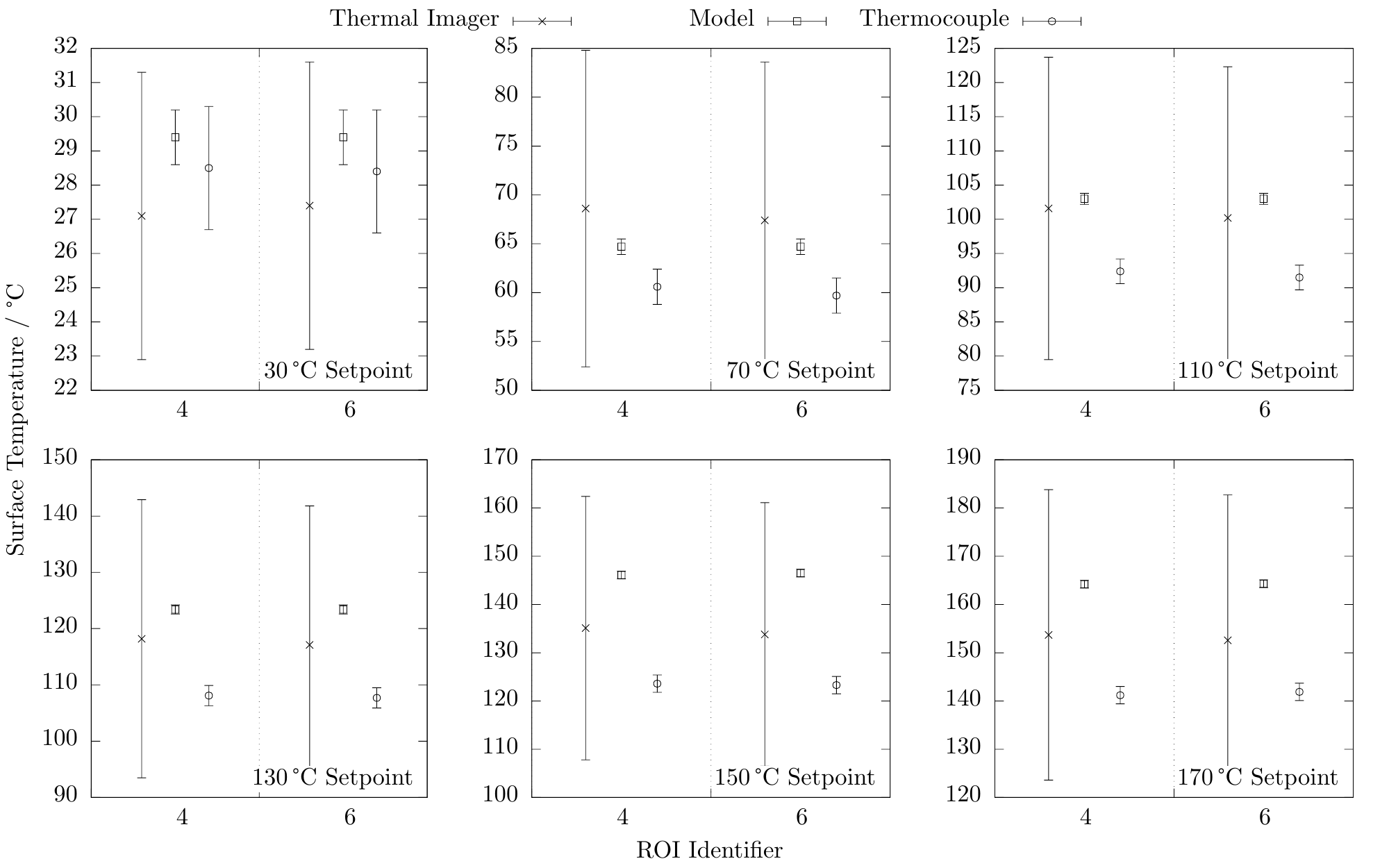}
\caption{A comparison between the surface temperature determinations of the uncoated regions using the thermal imager, thermocouples and the model with their equivalent uncertainties (\(k = 2\)) at each temperature setpoint.}
\label{fig:simulant_subsurface_artefact_test_plan_uncertainty_comparison}
\end{figure*}

The detailed analysis into the agreement between thermocouple, radiance and model temperature demonstrated a clear conclusion to the plate measurements. It was observed that as the plate heated, it began to physically warp and due to the fixed anchor points (the clamps) either side, the target plate may have flexed in the centre. This leads to an omission of the central column of regions of interest (ROI 2, ROI 5 and ROI 8) for the analysis. The coated regions of interest are denoted one through nine in reading order from the top-left corner. Additionally the thermocouples were spatially located (as seen in Fig.~\ref{fig:results_full_plate}) along the centre of the channels and so only those radiance temperature ROIs were considered: i.e. ROI 4 and ROI 6. A reduced comparison of this data can be seen in Fig.~\ref{fig:simulant_surface_artefact_test_plan_uncertainty_comparison} and Fig.~\ref{fig:simulant_subsurface_artefact_test_plan_uncertainty_comparison}.

\begin{figure}[h]
\centering
\includegraphics[width=0.5\textwidth,keepaspectratio]{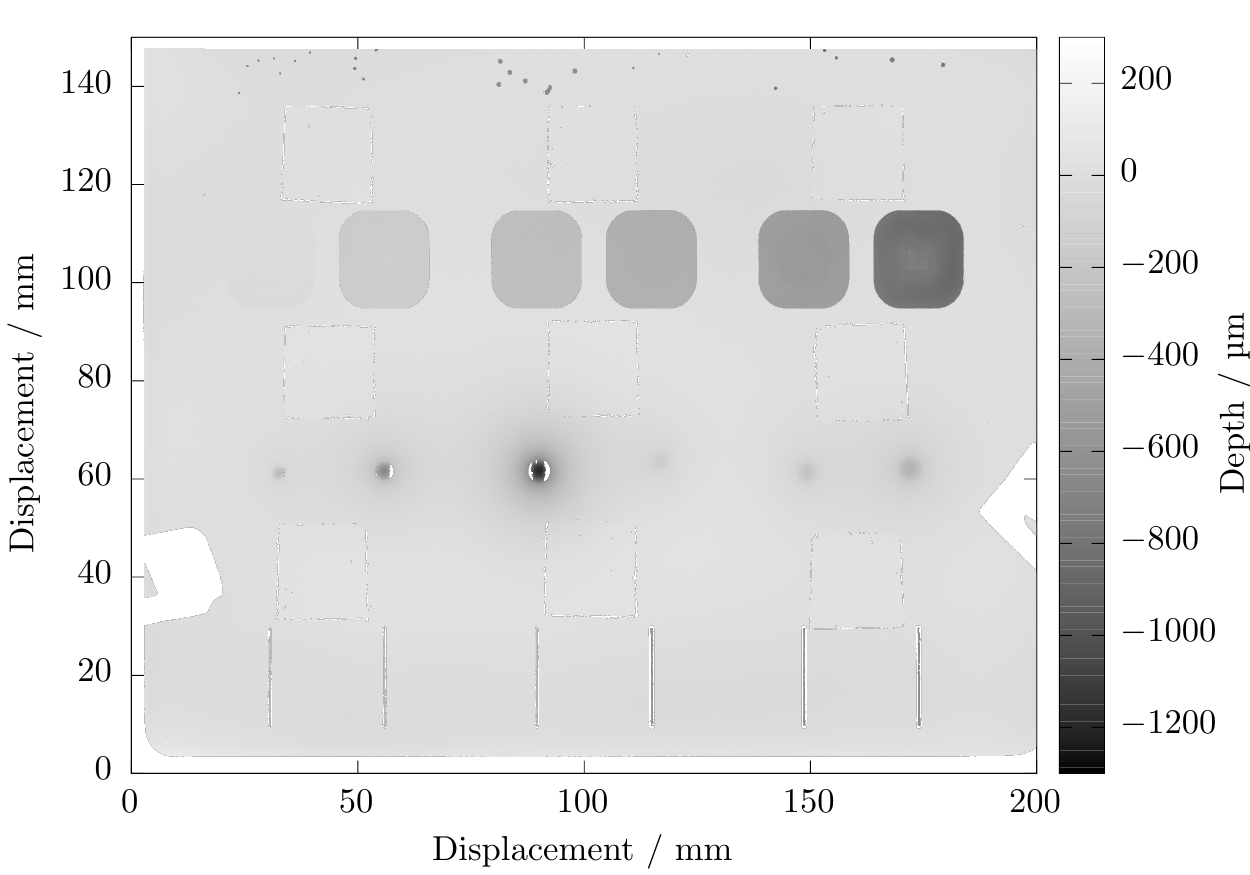}
\caption{Dimensional topography measurement of the plate and the surface artefacts. Note that regions of large depth gradient were not measurable and were out of range, these are white in this representation. A number of the pitting artefacts are not displayed here but were measured in a separate data set.}
\label{fig:simulant_topography}
\end{figure}

Surface topography characterisation of the plate was carried out using the equipment introduced in Section~\ref{subsec:dimensional_characterisation}. The complete measured topography can be seen in Fig.~\ref{fig:simulant_topography}. The widths and depths of each surface feature were determined using the topography data and the results are tabulated in full in Section~\ref{subsec:topography_characterisation}.

The effectiveness of the geometrical defect measurements using thermal imagers can be seen in Fig.~\ref{fig:simulant_defect_detected}, the complete dataset can be seen in the Appendix. It can be seen that for most temperature setpoints, the thermal imager is capable of successfully identifying both scratches and pitting on the surface of the plate. However, it should be noted that for lower temperature surfaces, there were a significantly larger number of false positives due to the higher level of noise present in the images. This indicates that thermal imagers could be used for defect detection in regimes where a sufficient level of thermal contrast is visible.

\begin{figure}[h]
\centering
\includegraphics[width=0.5\textwidth,keepaspectratio]{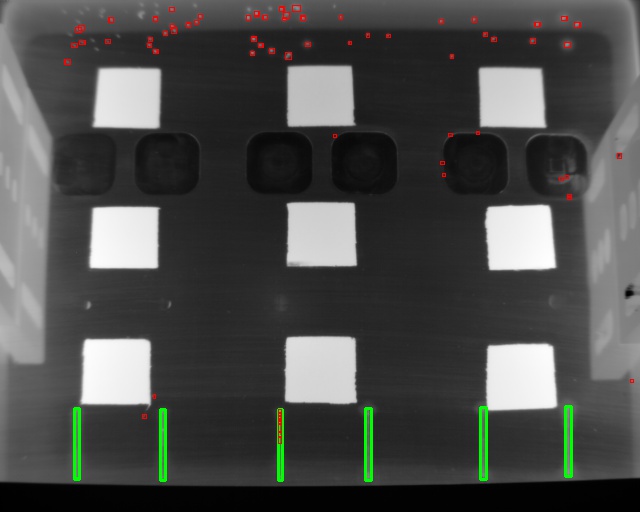}
\caption{Plate surface artefacts identified using the contrast-based image analysis. The results at \SI{170}{\celsius} and \SI{50}{\micro\second} show identified pits in red circles and scratches in green rectangles.}
\label{fig:simulant_defect_detected}
\end{figure}


\section{\bf Discussion} \label{sec:discussion}

In Fig.~\ref{fig:simulant_surface_artefact_test_plan_uncertainty_comparison} it can be seen for the surface plate measurements (Senotherm coated regions) that there is agreement between the thermocouple and thermal imager at all temperatures within the prescribed uncertainty. Aside from the \SI{70}{\celsius} setpoint, the data suggests the radiance temperature is too low; therefore the emissivity correction applied is not sufficient and a lower emissivity of \num{0.80} (compared to \num{0.85}) may be more suitable. Additionally, the model temperature is consistently greater than both measurements, which supports the postulation that the model does not fully account for all of the thermal losses from the experimental system.

For the sub-surface plate measurements (uncoated stainless-steel) seen in Fig.~\ref{fig:simulant_subsurface_artefact_test_plan_uncertainty_comparison}, the thermocouple and radiance measurements are in agreement at all temperature setpoints. In this case the radiance temperature is typically greater than the contact measurement; therefore a greater emissivity of \num{0.25} (as opposed to \num{0.20}) may be more appropriate. The model temperature does not agree with the thermocouples but does agree with the thermal imager. Both of these reduced dataset comparisons support a moderate reduction in the measurement uncertainty of each instrument, assuming the agreement with the model can be demonstrated.

The uncertainty budgets for the surface temperature determination of high emissivity regions detail the measurement uncertainty reaching \SI{6.8}{\celsius} (\(k = 2\)) at \SI{170}{\celsius}. An estimated maximum surface temperature for SNM containers is \SI{70}{\celsius} and so the equivalent measurement uncertainty for the coated surface using the thermal imager was \SI{3.4}{\celsius} (\(k = 2\)).

\begin{figure}[h]
\centering
\includegraphics[width=0.5\textwidth,keepaspectratio]{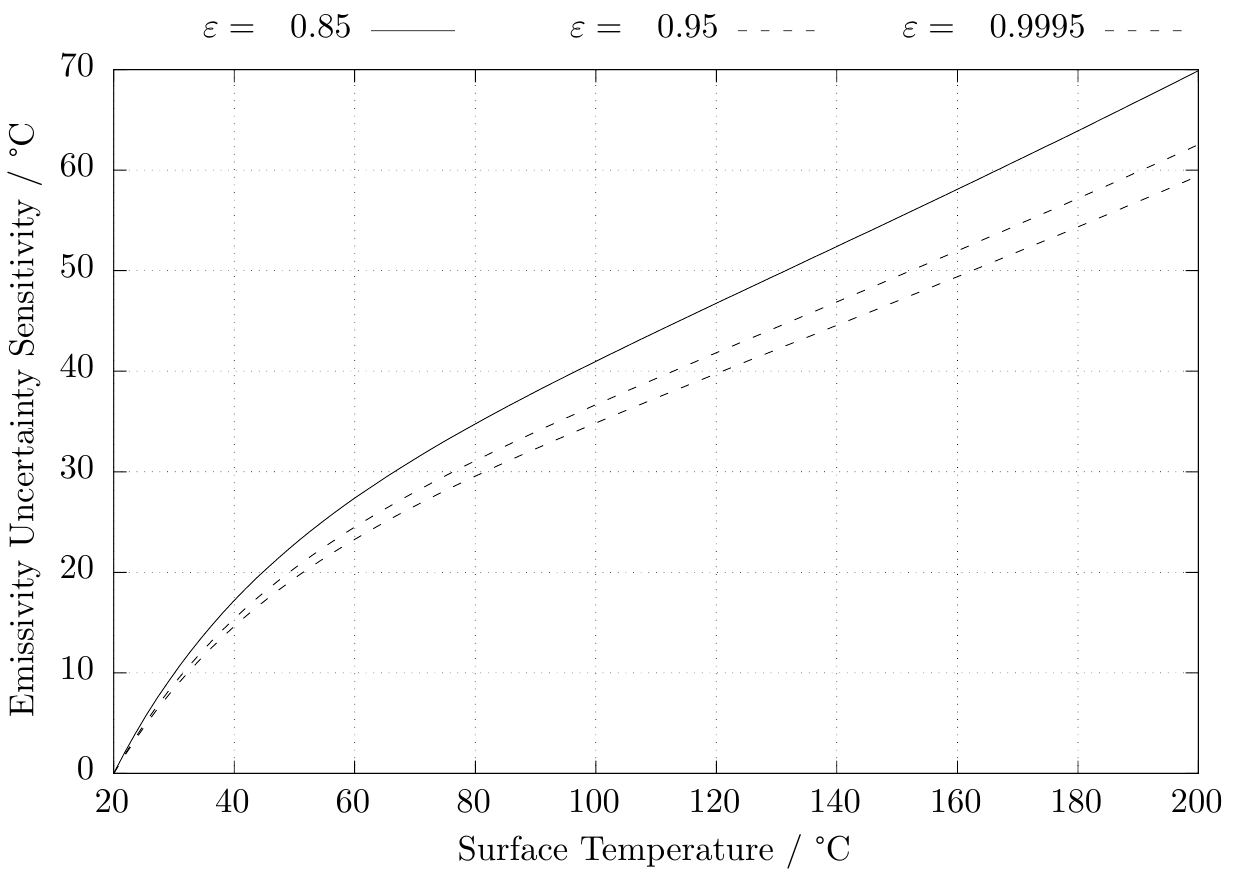}
\caption{Functional form of the emissivity sensitivity (Eq.~\ref{eq:emissivity_sensitivity}) for increasing temperature at a number of emissivities for a single wavelength and ambient temperature. Note that this sensitivity is largely dominated by the difference between apparent and ambient temperatures and that the emissivity itself has a smaller effect.}
\label{fig:simulant_surface_emissivity_sensitivity}
\end{figure}

The intrinsic thermal imager uncertainty is nominally consistent at \SI{0.3}{\celsius} (\(k = 1\)) across this temperature range and as seen from Tab.~\ref{tab:uncertainty}, the greatest source of uncertainty for temperature determination is from the emissivity measurement. The estimation for emissivity uncertainty of approximately \num{0.05} is close to national measurement institute grade measurements (as good as \num{0.01} (\(k = 1\)) depending on material) and so this particular component would reduce to a standard uncertainty of \SI{0.5}{\celsius}. Further reduction of this component requires reducing the sensitivity component (Eq.~\ref{eq:emissivity_sensitivity}) which is dominated by the disparity between apparent and ambient temperatures as opposed to the emissivity itself (see Fig.~\ref{fig:simulant_surface_emissivity_sensitivity}).

It should be noted that these equations are dependent on the spectral response of the instrument in use and the ambient temperature. Throughout these calculations a value of \SI{3.85}{\micro\metre} has been used alongside a background temperature of \SI{20}{\celsius}, but if a long-wave infrared instrument would be considered, then that would result in greater sensitivity coefficients throughout. 

\section{\bf Conclusion} \label{sec:conclusion}

The primary objectives of this study were to investigate and evaluate the feasibility of using a thermal imager to measure surface temperature -- both to monitor container activity and to characterise surface and sub-surface defects to be deployed for ongoing container monitoring and management. The primary activities undertaken in this project were the design and manufacture, followed by temperature and dimensional evaluation of a plate. The principle conclusion to be drawn from this study was that combined defect detection and temperature measurement of containers with thermal imaging is feasible. Surface temperature determination agreed with contact thermometers within the estimated measurement uncertainties. The smallest \SI{0.50}{\milli\metre} diameter pits and \SI{0.8}{\milli\metre} wide and \SI{0.26}{\milli\metre} deep scratches were successfully detected. These each place conservative estimates on the limits of detection for the method. However, a number of challenges were still encountered and revisions to both the hardware and measurement capture and analysis procedure would be recommended.

The largest issue with the hardware was the use of the clamps to pin either side of the target plate. The assembly could be redesigned to accommodate an internal ring of identical material to the housing that uniformly disperses pressure from the target plate to the thermal conductor plate. This top-edge of the housing may then be recessed to permit observation at multiple angles. Deployment of additional thermometers in each of the channels would have provided greater insight into the lateral uniformity of the assembly.

Emissivity had the largest impact on the uncertainty budget. The poor understanding of the particular surface emissivities resulted in a large discrepancy between thermocouple and radiance temperatures; whilst the large uncertainty component caused an extremely large impact on the standard uncertainty for radiance temperature. As noted, this component will be improved through traceable emissivity measurement, however will still be affected greatly by surface and ambient temperatures, instrument spectral response and potential application departures from the assumptions specified in the Appendix. Obstructions to deployment not explored here include radiation hardening of instrumentation, device control and management and instrument field of view optimisation.

Measurements during the sub-surface campaign demonstrated that it was not possible to identify the defects using passive thermal imaging. It may be possible to detect these regions during transient thermal excursions in the laboratory but deployment to in store containers would require a combined observation and illumination system.

The detection algorithm used in this study was a contrast-based technique that is invariant to absolute surface temperature and is enabled by apparent radiance variation across the surface. The strength of this method was that it was possible to identify artefacts with effective emissivity enhancements (i.e. pits and scratches); however, it was limited by the classification criteria. Further image analysis methods include developing a customised Haar-Cascade classifier model and integration into an AdaBoost implementation. These methods could be supplemented by the existing manual feature detection carried out to build a training data set, to then develop a multi-modal analytics model incorporating visual and thermal patterns as well as utilising the temperature measurement itself.


\bibliographystyle{unsrt}
\bibliography{thermal_and_geometrical_evaluation_of_a_test_artefact}

\section{\bf Appendix} \label{sec:appendix}
\subsection{Temperature determination sensitivity coefficients} \label{subsec:sensitivity}

Sensitivity coefficients for the three parameters: apparent radiance temperature \(T_{app}\), emissivity \(\varepsilon\) and ambient temperature \(T_{amb}\) with respect to the surface temperature \(T_b\) were determined using a partial derivative analysis.

Radiance from a surface \(L_{\lambda, b}\) is defined by

\begin{equation}
		L_{\lambda , b} = \frac{c_1}{\lambda^5} \frac{1}{e^{\nicefrac{c_2}{\lambda T_b}}-1} \mathrm{.}
		\label{eq:radiance}
\end{equation}

Where \(\lambda\) is the wavelength, \(c_1\) is the first radiation constant and \(c_2\) is the second radiation constant \cite{ref:thermal_radiation_heat_transfer}.

A simple equivalence between the apparent radiance \(L_{app}\), surface emissivity, radiance \(L_{\lambda, b}\) and ambient radiance \(L_{amb}\) under the assumption of single surface reflections is given by

\begin{equation}
		L_{app} = \varepsilon L_{\lambda,b} + \left( 1 - \varepsilon \right) L_{amb} \mathrm{.}
		\label{eq:apparent_radiance}
\end{equation}

Solving Eq.~\ref{eq:apparent_radiance} for the surface temperature gives

\begin{equation}
		T_b = \frac{c_2}{\lambda } ln\left[ \varepsilon \left[ \frac{1}{e^{\nicefrac{c_2}{\lambda T_{app}}}-1} -  \frac{\left( 1 - \varepsilon \right) }{e^{\nicefrac{c_2}{\lambda T_{amb}}}-1} \right]^{-1} + 1 \right]^{-1} \mathrm{.}
		\label{eq:surface_temperature}
\end{equation}

Eq.~\ref{eq:surface_temperature} was used throughout this project to determine the surface temperature using the measured temperatures, the estimated spectral mid-point (\SI{3.85}{\micro\metre}) and estimated emissivity. It should be noted that usage of this should consider: potential impact from multiple reflections, how to assess the spectral dependency (integrate over the specific spectral range of the system and optics used), the emissivity value used. In particular, the total hemispherical emissivity for a surface was used here, however a detailed assessment would consider the directional and spectral emissivity dependence in this evaluation.

The partial derivative of Eq.~\ref{eq:surface_temperature} with respect to the apparent radiance temperature, ambient temperature and surface emissivity are given below.

\begin{align}
		\begin{aligned}
				\frac{\partial T_b}{\partial T_{app}} =& \left[ \frac{c_2}{\lambda T_{app}} \right]^2 \left[ \varepsilon e^{\nicefrac{c_2}{\lambda T_{app}}} \left( e^{\nicefrac{c_2}{\lambda T_{amb}}} - 1 \right)^2 \right] \cdot \\
				& \left[ \varepsilon  e^{\frac{c_2}{\lambda} \left( \frac{1}{T_{app}} + \frac{1}{T_{amb}} \right)} + \left( \varepsilon - 1 \right) e^{\nicefrac{c_2}{\lambda T_{amb}}} + e^{\nicefrac{c_2}{\lambda T_{app}}}  \right]^{-1} \cdot \\
				& \left[ \varepsilon \left( e^{\nicefrac{c_2}{\lambda T_{app}}} - 1  \right)  -  e^{\nicefrac{c_2}{\lambda T_{app}}}  +  e^{\nicefrac{c_2}{\lambda T_{amb}}}  \right]^{-1} \cdot \\
				& ln\left[ 1 + \frac{\varepsilon}{ \frac{1}{ e^{\nicefrac{c_2}{\lambda T_{app}}} - 1 } + \frac{\varepsilon - 1}{ e^{\nicefrac{c_2}{\lambda T_{amb}}} - 1 } } \right]^{-2} 
		\label{eq:apparent_temperature_sensitivity}
		\end{aligned}
\end{align}

\begin{align}
		\begin{aligned}
				\frac{\partial T_b}{\partial T_{amb}} =& \left[ \frac{c_2}{\lambda T_{amb}} \right]^2 \left[ \varepsilon \left( \varepsilon - 1 \right) e^{\nicefrac{c_2}{\lambda T_{amb}}} \left( e^{\nicefrac{c_2}{\lambda T_{app}}} - 1 \right)^2 \right] \cdot \\
				& \left[ \varepsilon  e^{\frac{c_2}{\lambda} \left( \frac{1}{T_{app}} + \frac{1}{T_{amb}} \right)} + \left( \varepsilon - 1 \right) e^{\nicefrac{c_2}{\lambda T_{amb}}} + e^{\nicefrac{c_2}{\lambda T_{app}}}  \right]^{-1} \cdot \\
				& \left[ \varepsilon \left( e^{\nicefrac{c_2}{\lambda T_{app}}} - 1  \right)  -  e^{\nicefrac{c_2}{\lambda T_{app}}}  +  e^{\nicefrac{c_2}{\lambda T_{amb}}}  \right]^{-1} \cdot \\
				& ln\left[ 1 + \frac{\varepsilon}{ \frac{1}{ e^{\nicefrac{c_2}{\lambda T_{app}}} - 1 } + \frac{\varepsilon - 1}{ e^{\nicefrac{c_2}{\lambda T_{amb}}} - 1 } } \right]^{-2} 
		\label{eq:ambient_temperature_sensitivity}
		\end{aligned}
\end{align}

\begin{align}
		\begin{aligned}
				\frac{\partial T_b}{\partial \varepsilon} =& \frac{c_2}{\lambda} \left[ \frac{1}{e^{\nicefrac{c_2}{\lambda T_{amb}}} - 1} - \frac{1}{e^{\nicefrac{c_2}{\lambda T_{app}}} - 1} \right] \cdot \\
				& \left[ \frac{1}{ e^{\nicefrac{c_2}{\lambda T_{app}}} - 1 } + \frac{\varepsilon - 1}{ e^{\nicefrac{c_2}{\lambda T_{amb}}} - 1 } \right]^{-1} \cdot \\
				& \left[ \varepsilon + \frac{1}{ e^{\nicefrac{c_2}{\lambda T_{app}}} - 1 } + \frac{\varepsilon - 1}{ e^{\nicefrac{c_2}{\lambda T_{amb}}} - 1 } \right]^{-1} \cdot \\
				& ln\left[ 1 + \frac{\varepsilon}{ \frac{1}{ e^{\nicefrac{c_2}{\lambda T_{app}}} - 1 } + \frac{\varepsilon - 1}{ e^{\nicefrac{c_2}{\lambda T_{amb}}} - 1 } } \right]^{-2} 
		\label{eq:emissivity_sensitivity}
		\end{aligned}
\end{align}

\subsection{Defect detection results} \label{subsec:defect_detection_results}
\begin{figure*}[h]
\centering
\subcaptionbox{Temperatures from \SIrange{30}{110}{\celsius}. \label{fig:defects_30_110}}
{\includegraphics[width=0.57\textwidth,keepaspectratio]{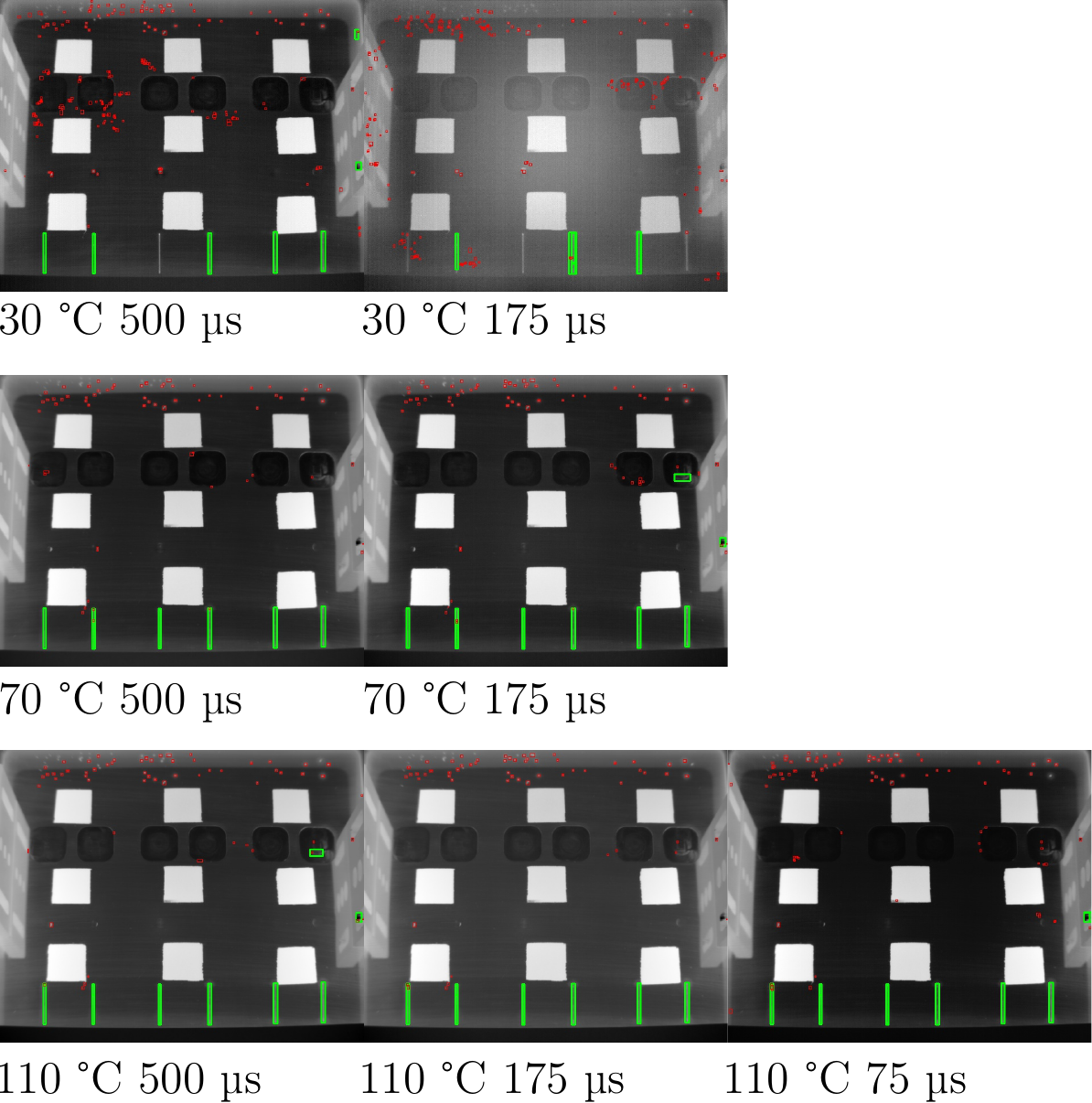}}
\vspace{\floatsep}
\subcaptionbox{Temperatures from \SIrange{130}{170}{\celsius}. \label{fig:defects_130_170}}
{\includegraphics[width=0.57\textwidth,keepaspectratio]{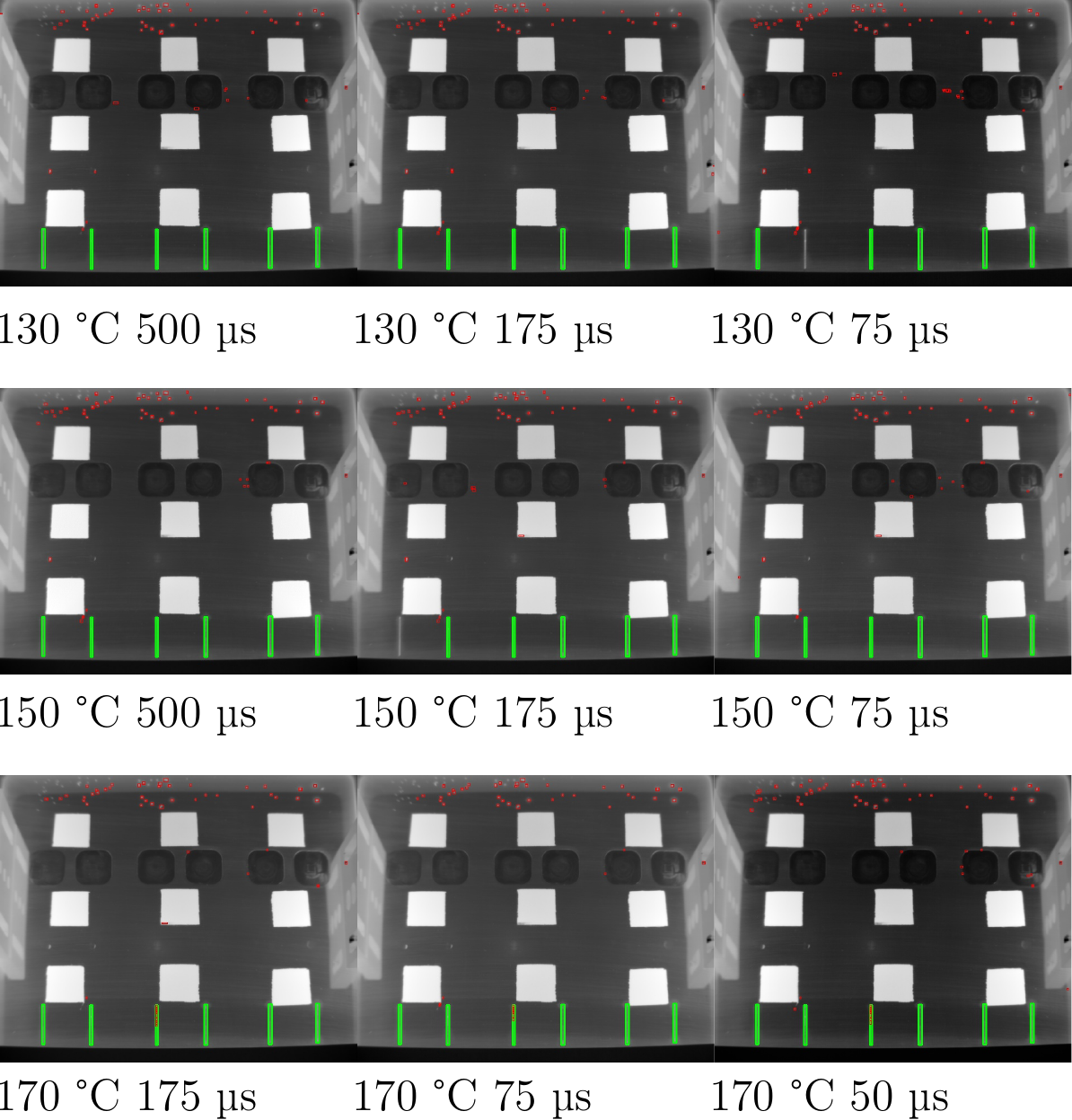}}
\caption{Geometric defects identified using the thermal during each of the measurement sets. The green outline corresponds to a detected scratch and the red outline denotes a pit.}
\label{fig:defect_detection}
\end{figure*}

A complete set of results from the defect detection can be seen in Fig.~\ref{fig:defects_30_110} and Fig.~\ref{fig:defects_130_170}; for each temperature setpoint the respective imager integration times are denoted. The green outline corresponds to a detected scratch and the red outline denotes a pit.

\subsection{Topography characterisation results} \label{subsec:topography_characterisation}
Results from the topography characterisation of the plate using a chromatic confocal point probe sensor can be seen in Tab.~\ref{tab:topography_results_pit} through Tab.~\ref{tab:topography_results_dent}. The artefact identification indicates the horizontal position from right to left in Fig.~\ref{fig:simulant_topography}.

\begin{table}[t]
		\centering 
		\caption{The mean depths and widths of pitting measured using scanning profilometry. The pitting group discriminates which of the six artefact zones each pit falls within, and each pit in a group was designed to be identical. \textsuperscript{\dag}This pit was not successfully measured and reported no depth.}
		\begin{tabular}{ M{3.0em} M{3.0em} M{3.0em} M{3.0em} M{3.0em} M{3.0em} }
		\Xhline{1.0pt}
		Pitting ID & Pitting Group & Mean Depth / \SI{}{\milli\metre} & \(u \left( k = 2 \right)\) & Width / \SI{}{\milli\metre} & \(u \left( k = 2 \right)\) \\
		\cmidrule(lr){1-6}
		1 & \multirow{2}{*}{1} & 0.798 & 0.008 & 1.30 & 0.13 \\
		2 &  & 0.745 & 0.008 & 1.20 & 0.13 \\
		\cmidrule(lr){1-6}
		3 & \multirow{3}{*}{2} & 0.826 & 0.033 & 0.80 & 0.13 \\
		4 &  & 0.820 & 0.009 & 0.80 & 0.13 \\
		5 &  & 0.777 & 0.026 & 0.80 & 0.13 \\
		\cmidrule(lr){1-6}
		6\textsuperscript{\dag} & \multirow{3}{*}{3} & - & - & 0.50 & 0.13 \\
		7 &  & 0.734 & 0.010 & 0.60 & 0.13 \\
		8 &  & 0.719 & 0.012 & 0.60 & 0.13 \\
		\cmidrule(lr){1-6}
		9 & \multirow{6}{*}{4} & 0.628 & 0.008 & 1.20 & 0.13 \\
		10 &  & 0.638 & 0.013 & 1.50 & 0.13 \\
		11 &  & 0.622 & 0.008 & 1.10 & 0.13 \\
		12 &  & 0.628 & 0.009 & 1.10 & 0.13 \\
		13 &  & 0.629 & 0.013 & 1.10 & 0.13 \\
		14 &  & 0.630 & 0.009 & 1.10 & 0.13 \\
		\cmidrule(lr){1-6}
		15 & \multirow{4}{*}{5} & 0.720 & 0.010 & 0.90 & 0.13 \\
		16 &  & 0.688 & 0.063 & 0.80 & 0.13 \\
		17 &  & 0.719 & 0.009 & 0.80 & 0.13 \\
		18 &  & 0.699 & 0.040 & 0.80 & 0.13 \\
		\cmidrule(lr){1-6}
		19 & \multirow{7}{*}{6} & 0.732 & 0.010 & 0.50 & 0.13 \\
		20 &  & 0.726 & 0.009 & 0.50 & 0.13 \\
		21 &  & 0.725 & 0.009 & 0.50 & 0.13 \\
		22 &  & 0.726 & 0.010 & 0.50 & 0.13 \\
		23 &  & 0.718 & 0.015 & 0.50 & 0.13 \\
		24 &  & 0.725 & 0.010 & 0.50 & 0.13 \\
		25 &  & 0.702 & 0.014 & 0.50 & 0.13 \\
		\Xhline{1.0pt}
		\end{tabular} 
		\label{tab:topography_results_pit}
\end{table}

\begin{table}[t]
		\centering 
		\caption{The mean depths and widths of each scratch.}
		\begin{tabular}{ M{3.0em} M{3.0em} M{3.0em} M{3.0em} M{3.0em} }
		\Xhline{1.0pt}
		Scratch ID & Mean Depth / \SI{}{\milli\metre} & \(u \left( k = 2 \right)\) & Width / \SI{}{\milli\metre} & \(u \left( k = 2 \right)\) \\
		\cmidrule(lr){1-5}
		1 & 0.62 & 0.01 & 1.3 & 0.2 \\
		2 & 0.60 & 0.01 & 1.3 & 0.2 \\
		3 & 0.58 & 0.01 & 1.2 & 0.2 \\
		4 & 0.26 & 0.01 & 0.8 & 0.2 \\
		5 & 0.26 & 0.01 & 0.8 & 0.2 \\
		6 & 0.26 & 0.01 & 0.8 & 0.2 \\
		\Xhline{1.0pt}
		\end{tabular} 
		\label{tab:topography_results_scratch}

		\vspace{1em}

		\centering 
		\caption{The mean depths and widths of each thinned region.}
		\begin{tabular}{ M{3.0em} M{3.0em} M{3.0em} M{3.0em} M{3.0em} }
		\Xhline{1.0pt}
		Thinning ID & Mean Depth / \SI{}{\milli\metre} & \(u \left( k = 2 \right)\) & Width / \SI{}{\milli\metre} & \(u \left( k = 2 \right)\) \\
		\cmidrule(lr){1-5}
		1 & 0.83 & 0.01 & 20.2 & 0.2 \\
		2 & 0.52 & 0.01 & 20.2 & 0.2 \\
		3 & 0.39 & 0.01 & 20.1 & 0.2 \\
		4 & 0.28 & 0.01 & 20.0 & 0.2 \\
		5 & 0.16 & 0.01 & 20.0 & 0.2 \\
		6 & 0.06 & 0.01 & 19.9 & 0.2 \\
		\Xhline{1.0pt}
		\end{tabular} 
		\label{tab:topography_results_thinning}

		\vspace{1em}

		\centering 
		\caption{The maximum depth and width of each dent.}
		\begin{tabular}{ M{3.0em} M{3.0em} M{3.0em} M{3.0em} M{3.0em} }
		\Xhline{1.0pt}
		Dent ID & Max Depth / \SI{}{\milli\metre} & \(u \left( k = 2 \right)\) & Width / \SI{}{\milli\metre} & \(u \left( k = 2 \right)\) \\
		\cmidrule(lr){1-5}
		1 & 0.36 & 0.02 & 5.5 & 0.1 \\
		2 & 0.24 & 0.02 & 4.8 & 0.1 \\
		3 & 0.20 & 0.02 & 5.7 & 0.1 \\
		4 & 1.20 & 0.02 & 4.7 & 0.1 \\
		5 & 0.68 & 0.02 & 3.6 & 0.1 \\
		6 & 0.32 & 0.02 & 2.5 & 0.1 \\
		\Xhline{1.0pt}
		\end{tabular} 
		\label{tab:topography_results_dent}
\end{table}

\end{document}